\documentclass{aa}
\usepackage{graphicx}
\input{psfig}
\usepackage{natbib}
\bibpunct{(}{)}{;}{a}{}{,} 

\usepackage{times}
\newcommand{\twCO}  {\mbox{$^{12}$CO}}       
\newcommand{\Jone}{\mbox{$J$=1--0}}
\newcommand{\Jtwo}{\mbox{$J$=2--1}}

\def\zper{\mbox{$\zeta~\mathrm{Per}$}}
\defcitealias{boisseal05}{B05}

\def\kms{\mbox{{km~s}$^{-1}$}}
\def\units#1#2{\mbox{{#1}$^{#2}$}}
\def\cmd{\units{cm}{-2}}
\def\cmt{\units{cm}{-3}}

\def\CHP{\mbox{{\rm CH}$^+$}}

\def\CHl{CH~$\lambda$4300}
\def\WCHl{$W_{4300}$(CH)}
\def\CHPlb{CH$^{+}$~$\lambda$4232}
\def\CHPla{CH$^{+}$~$\lambda$3957}
\def\WCHPla{$W_{3957}$(CH$^+$)}
\def\WCHPlb{$W_{4232}$(CH$^+$)}
\def\HI{\ion{H}{i}}
\def\H2{H$_2$}

\def\co{$^{12}$CO(1-0)}
\def\co21{\mbox{$^{12}\mathrm{CO}(2-1)$}}
\def\vlsr{\mbox{$V_{\mathrm{LSR}}$}}

\def\Sec#1{Sect.~\ref{s:#1}}

\def\Tab#1{{Table~\ref{t:#1}}}        
\def\Fig#1{{Fig.~\ref{f:#1}}}    

\begin{document}

\title{CO emission and variable CH and CH$^+$ absorption
towards HD~34078: evidence for a nascent bow shock ?
\thanks{Based on observations made mainly at IRAM, Observatoire de Haute Provence 
(France), McDonald Observatory (USA) and with FUSE}}

\titlerunning{CO emission and variable CH and CH$^+$ absorption  
towards HD~34078}

\author{P. Boiss\'e\inst{1}, E. Rollinde\inst{2}, P. Hily-Blant\inst{3}, 
J. Pety\inst{4}, S.R. Federman\inst{5},
Y. Sheffer\inst{5}, G. Pineau des For\^ets\inst{6}, E. Roueff\inst{7}, 
B-G Andersson\inst{8},
G. H\'ebrard\inst{2}}
\institute{Institut d'Astrophysique de Paris (IAP),  UMR7095 CNRS, 
           Universit\'e Pierre et Marie Curie-Paris6, 98 bis boulevard
           Arago, 75014 Paris, France \\
           \email{boisse@iap.fr}
       \and
            Institut d'Astrophysique de Paris, UMR7095 CNRS, 
	    Universit\'e Pierre et Marie Curie-Paris6, 98 bis boulevard Arago,
            75014 Paris, France
        \and
           IRAM, Domaine Universitaire, 300 rue de la Piscine, 38406
	   Saint-Martin-d'H\`eres, France ; Laboratoire d'Astrophysique,
	   Observatoire de Grenoble, BP 53, 38041 Grenoble Cedex 9,
	   France 
         \and
           Institut de Radioastronomie Millim\'etrique, 300 rue de la
	   Piscine, 38406 Saint Martin d'H\`eres, France ; Observatoire de Paris,
	   61 Av. de l'Observatoire, 75014 Paris, France
         \and
           Department of Physics and Astronomy, University of Toledo, Toledo, OH 43606, USA
 	 \and
           IAS, Universit\'e d'Orsay, 91405 Orsay Cedex, France
         \and 
           LUTH, Observatoire de Paris-Meudon, 92195 Meudon Cedex,
	   France
         \and 
            NASA Ames Research Center, Moffett Field, CA 94035, USA
         }

\authorrunning{P. Boiss\'e et al.}
\date{Accepted 2009 January 24. Received 2008 September 15.}

\abstract
{The runaway star HD~34078, initially selected to investigate small
scale structure in a foreground diffuse cloud 
has been shown to be surrounded by highly excited \H2, the origin of
which is unclear.}
{We first search for an association between the foreground
cloud and HD~34078. Second, we extend previous investigations of
temporal absorption line variations (CH, \CHP, \H2) 
}
{We have mapped the \co21\ emission at 12\arcsec\ resolution around
HD~34078's position, using the 30~m IRAM antenna. The follow-up of CH and 
\CHP\ absorption lines has been extended over 5 more years.
In parallel, CH absorption towards the reddened star \zper\ 
have been monitored to check the
homogeneity of our measurements. Three more FUSE spectra have been
obtained to search for $N$(\H2) variations.}
{CO observations show a pronounced maximum near HD~34078's position, 
clearly indicating that the star and diffuse cloud are associated. The
optical spectra confirm the reality of strong, rapid and
correlated CH and \CHP\ fluctuations. 
On the other hand, $N$(\H2, $J=0$) has varied by less than
5\% over 4 years, indicating the absence of marked 
density structure at scales below 100 AU. We also discard $N$(CH)
variations towards \zper\ at scales less than 20 AU.}
{Observational constraints from this work and from 24~$\mu$m dust
emission appear to be consistent with \H2\ excitation but
inconsistent with steady-state bow shock models and rather suggest 
that the shell of compressed gas surrounding HD~34078
 or lying at the boundary of a small foreground clump, 
is seen at an early stage of the interaction. The CH and \CHP\ time
variations as well as their large abundances are likely due to chemical
structure in the shocked gas layer located at the stellar wind/ambient
cloud interface. Finally, the lack of
variations for both $N$(\H2, $J=0$) towards HD~34078 and $N$(CH) towards
\zper\ suggests that quiescent molecular gas is not subject to pronounced 
small-scale structure.}


\maketitle

\section{Introduction}

During the past decade strong evidence has accumulated indicating that 
the spatial distribution of species like \ion{Na}{i} and \ion{Ca}{ii} 
within neutral interstellar (IS) gas displays significant 
structure at AU scales \citep{crawford03,lauroesch07,welty07,weltyal08}. 
\HI\ itself, within the 
cold neutral medium at least, shows such structure \citep{frailal94,heiles97,deshpande07,weisbergal07}. In 
diffuse molecular gas, similar conclusions have been reached for tracers 
like H$_2$CO, HCO$^+$, and OH  \citep{mooreal95,lisztal00}. 
Whether or not these spatial variations correspond to true density structure 
[i.e. to local fluctuations of $n$(H$_2$)] is obviously of key importance 
for the modelling of physical and chemical processes within molecular gas. 

To investigate this question, a time variation study 
of \H2, CH, and \CHP\ interstellar absorption lines towards the O9.5V 
runaway star AE Aur, HD~34078 has been undertaken by 
\citet[\ , hereafter R03]{rollindeal03} and \citet[\ , hereafter B05]{boisseal05}. 
This bright star
is significantly reddened [E(B-V) = 0.53] and its optical spectrum 
displays strong absorption lines from e.g. CH, \CHP, CN
\citep{federmanal94},
typical of diffuse molecular clouds. HD~34078 has a large proper motion 
of $\mu$ = 43 mas yr$^{-1}$, corresponding to a transverse velocity of 
103 \kms\ or 22 AU yr$^{-1}$ for a distance $D=530$~ pc (this value will be
used in the following for consistency with \citetalias{boisseal05}; it
is 
compatible with the trigonometric
parallax estimate of $446^{+220}_{-111}$~pc). The line of sight is thus drifting 
rapidly through the foreground gas; successive column density 
measurements then provide a ``cut'' through the cloud revealing its spatial 
density structure over scales which typically range from 1 to 100 AU for 
time separations ranging from a few weeks to a few years.

The five first FUSE spectra discussed in B05 showed that highly excited \H2\
gas is present along the line of sight, together with more standard quiescent 
gas at $T\simeq$ 80 K. The presence of significant amounts of \H2\ with an 
excitation energy higher that 2500 K (corresponding to the $v=0, J=5$ state) 
is very rare \citep[another remarkable case is that of HD~37903 studied by][]{meyeral01}. 
This is certainly related to the fact that by chance, along its long 
path from 
the Orion nebula where it was ejected a few millions years ago 
\citep{hoogerwerfal01}, HD~34078 has 
recently encountered a dense interstellar cloud with which it is currently 
interacting, leading to the present-day apparence of the IC~405 nebula 
\citep{herbig58}.

In previous modelling work of the properties of the gas along the line
of sight (B05), it was found that two components are required to account for the observed 
absorption lines:

- highly excited \H2\ located in a bow shock, where the stellar wind 
impacts onto the ambient medium. Illumination of this 
gas by the strong UV field of the star satisfactorily explains the 
observed \H2 excitation diagram. The presence of such a bow shock around 
HD~34078 was suspected on the basis of IRAS data \citep{vanburenal95} 
and has been confirmed recently by higher spatial resolution 
$Spitzer$ observations \citep{franceal07},

- a foreground quiescent cloud (supposedly unrelated to HD~34078/IC~405) 
giving rise to absorption from cold \H2\ and other species typical of 
translucent material (CH, CO, CN).

However, in this scenario, the close similarity of the radial velocity of both
components appears as a pure coincidence. Further, the CH/\H2\ ratio towards 
HD~34078 displays an anomalously high value, about three to four times larger 
than the one commonly found for other lines of sight. 
This led us to suspect that HD~34078 might be more closely related 
to the cold cloud than assumed in the above model, even if at first sight,
this appears difficult to reconcile with the low \H2 temperature ($T$ = 77 K) 
derived from the population of the $J$ = 0 and $J$ = 1 levels and the presence 
of a variety of molecules that HD~34078 might easily photodissociate.

The potential relation between HD~34078 and the cloud probed 
has important implications regarding the investigation of small-scale
structure. Indeed, in the case of a real association, any mechanical or 
radiative interaction might 
significantly affect the initial structure. Further, the 
gas flow in the bow shock may be subject to instabilities which 
could lead to time variations of a different nature, not necessarily 
associated with spatial structure. 

Thus, one objective of the present 
study is to clarify the relation between 
HD~34078 and the cold cloud. To this aim, we have undertaken 
observations of CO emission in the field surrounding the O star. 
The second goal of our work is to extend the search for absorption line 
variations in ground-based spectra (CH and \CHP\ mainly) and in FUSE 
spectra (\H2) by adding recent data obtained after the initial 
studies by R03 and B05 were completed. 

This paper is organised as follows. We first present high spatial resolution 
\co21\ and $^{12}$CO(1-0) observations performed at the 30~m IRAM telescope and their 
implications concerning the location and properties of the quiescent
component (Sect. 2).
In Sect. 3, we analyse a new series of optical spectra which allow us to 
study in detail the variations of CH and \CHP\ column densities 
[hereafter $N$(CH) and $N$(\CHP)] and velocity profiles. 
Time variations of \H2\ column densities [hereafter
$N$(\H2)] are then discussed in the light of the observed CH and \CHP\ 
variations (Sect. 4). In Sect. 5, we summarize the main observational
results for \zper\ and HD~34078, in particular for readers not interested in
details concerning observations. Then, we discuss the implications
of our observations in terms of processes related to the interaction 
between HD~34078 and the surrounding gas and small-scale structure
in foreground unperturbed gas (Sect. 6). Finally, we summarize our 
conclusions and present some prospects concerning possible observational
signatures of the future evolution of HD~34078's close environment.

\section{CO emission towards HD~34078}

$^{12}$CO emission can be used to investigate a possible 
connection 
between HD~34078 and the molecules seen in absorption. Indeed, 
in the two-component model presented in B05, CO molecules 
(as well as CH, \CHP\ etc.) lie far in front of HD~34078 and 
if this picture were correct, the morphology of the CO emission should 
not correlate in any manner with the star position. Otherwise, 
any relation between the CO emission map and HD~34078's position would 
be a clear indication that the star is closely related to the foreground 
cloud.

\subsection{IRAM-30m observations}

Emission of the rotational transitions of $^{12}$CO was
observed at the IRAM-30m telescope during three consecutive
nights on February 11, 12 and 13, 2004. 
Using the HERA
$3\times3$ multi-beam array \citep{schusteral04}, we mapped
the $^{12}$CO(2-1) emission in a $66 \times 66$~arcsec$^2$
region centred on HD~34078's position. The single side-band
receiver temperature was in the range 120 to
180~K. Observations were done under good weather conditions
with 2~mm water vapor and a zenith sky opacity at 230~GHz 
$\tau_{230}=0.15$, resulting in system temperatures in the
range 250--350~K. Chopper-wheel calibration was done every
10--15 minutes. Pointing was checked frequently, ensuring an
accuracy of 2\arcsec. We used the VESPA autocorrelator as a
spectrometer covering 160~\kms\ with a resolution power of
$7\times10^{5}$ or $\delta v=0.4$~\kms\ ($\delta \nu =
0.3$~MHz; see \Tab{CO} for CO observation parameter values).  
The map was done in raster mode to allow deep
integration of about 40~min. to ensure a rms of 15~mK
(antenna temperature scale) in each velocity channel. It
consists in a regular grid of $12\times12$ positions
observed with a sampling of 6\arcsec, and the data are thus
only slightly undersampled ($HPBW=11.7\arcsec$\ at
230~GHz). A 5-point cross centred at offsets $(-3,3)$
with 12\arcsec\ steps was observed simultaneously with the
single-pixel receivers facility, 
in the $^{12}$CO(1-0) and (2-1) transitions ($HPBW=22\arcsec$ at 115~GHz) to derive
the excitation conditions of the gas.

\subsection{Results and analysis of the CO data}

\begin{figure}
\includegraphics[width=\hsize,angle=-90]{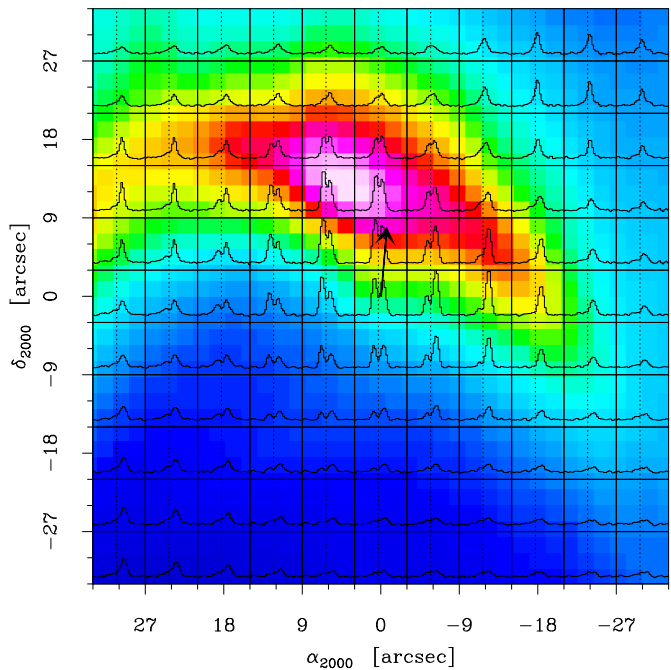}
\caption{$^{12}$CO(2-1) spectra overlaid on the MIPS 
24~$\mu$m emission of \cite{franceal07} (color scale; note that the
 IR map
 is saturated near the maximum at the (3,12) position). The
	 CO LSR velocity scale is in the range 0--13~\kms\ and the
	  antenna temperature scale ranges from $-$0.1 to
	  0.6~K. The observed positions correspond to the center of each panel;
 the dotted line indicates a
	  constant velocity $\vlsr=5.9$~\kms, associated
	  with the CH and CH$^+$ lines (see below). 
HD~34078 is located at offsets (0,0); 
	  the direction of its proper motion
	  is indicated by the arrow.}
\label{f:hera}
\end{figure}

The resulting map of the $^{12}$CO(2-1) emission is shown in
\Fig{hera}, in which the nearly fully-sampled spectra were projected on
a 6~\arcsec $\times$ 6~\arcsec\ grid centred on the (0,0) offset.
Thermal dust emission has also been observed recently by 
\citet{franceal07} and their MIPS 24~$\mu$m is overlaid by the CO
spectra.

\begin{figure}
\includegraphics[width=\hsize,angle=-90]{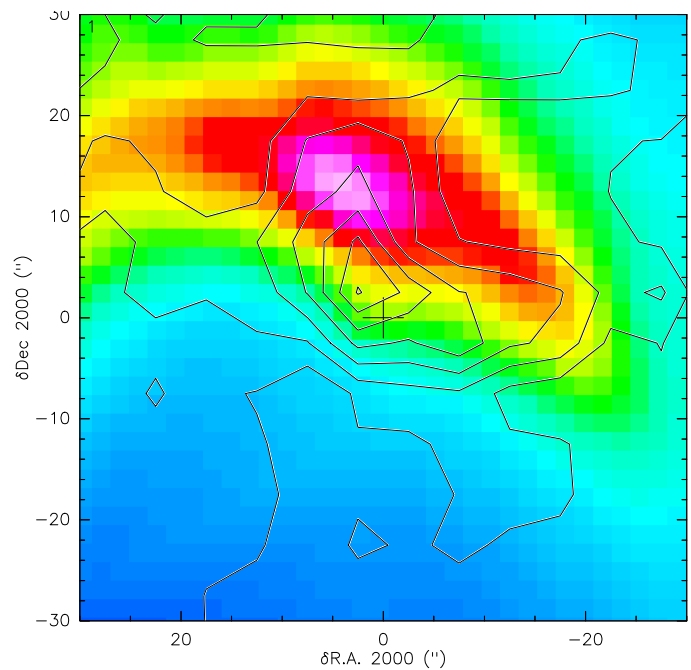}
\caption{The integrated $^{12}$CO(2-1) emission (contours) overlaid on the MIPS 
24~$\mu$m emission of \cite{franceal07} (color scale). The CO profiles
 have been integrated in the interval 2.5 - 10 \kms; the first contour
 level and spacing between successive contours is 0.189~K.\kms.
}
\label{f:integrated}
\end{figure}

\begin{figure}
\includegraphics[width=\hsize,angle=-90]{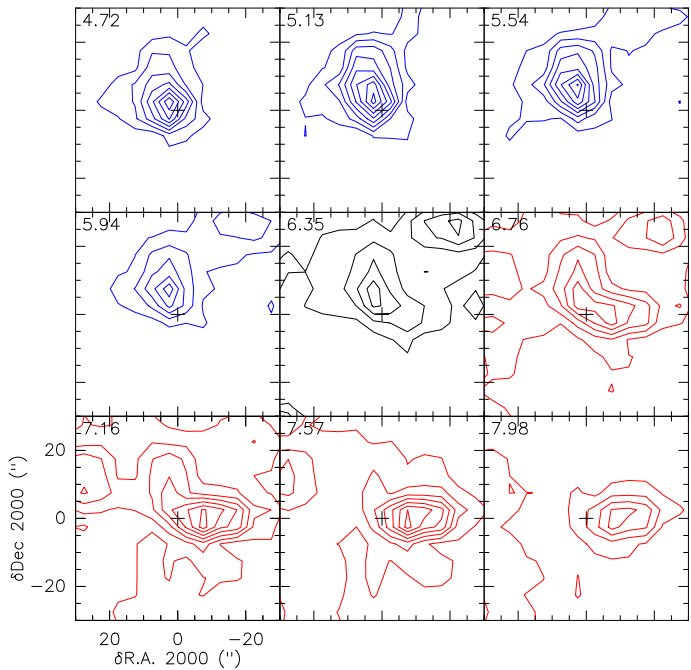}
\caption{Channel maps of the $^{12}$CO(2-1) emission averaged over an 
interval of 0.41~\kms (the value in the upper
 left corner is the center of the velocity interval considered). 
Each panel corresponds to the region shown in \Fig{hera}.
The first contour level and spacing between successive contours is 3.5~mK.
}
\label{f:channels}
\end{figure}

We also present a map of the integrated \co21\ intensity superimposed 
onto the MIPS 24~$\mu$m emission (\Fig{integrated}) as well as channel
maps (\Fig{channels}). 
The \co21\ integrated intensity is clearly stronger around the star 
position (0,0). 
While the NE and SW region are nearly devoid of emission, the region of
enhanced \co21\ emission is resolved and displays an elongated morphology
(in the NE-SW direction) and extent which are both relatively similar to those
of the brightest part of the IR arc. The \co21\
peak is apparently offset southward by about 9~\arcsec\ with 
respect to the 24~$\mu$m maximum but, given the accuracy of absolute
positions in the IR map ($\simeq$ 6~\arcsec; K. France, private communication)
and \co21\ emission ($\simeq$ 2~\arcsec), it is not clear whether this
offset is really significant.

The profiles are double peaked close to
the star (in an approximate circle of diameter 20-30~\arcsec; note the blue
and red components appearing on each side of the vertical dotted line at 
5.9~\kms\ plotted in \Fig{hera}). The spectra closest
to the star display a kind of mirror symmetry about an axis coincident with HD
34078's path, with the blue emission line stronger to the E and the red one
stronger to the W. This is apparent in the channel maps
(\Fig{channels}), where the red component peaks to the W while the
blue one peaks to the NE. The symmetry quickly disappears as the profiles 
become single
peaked further away. Averaging spectra in the NW or SE areas clearly shows
that weak wings are present there over the entire velocity range (\vlsr = 3
- 10~\kms) covered by the more intense emission seen in the central
part. The systemic velocity of the ambient molecular gas is 
\vlsr\ $\simeq$ 6.5~\kms.

\begin{figure}
\includegraphics[width=\hsize,angle=-90]{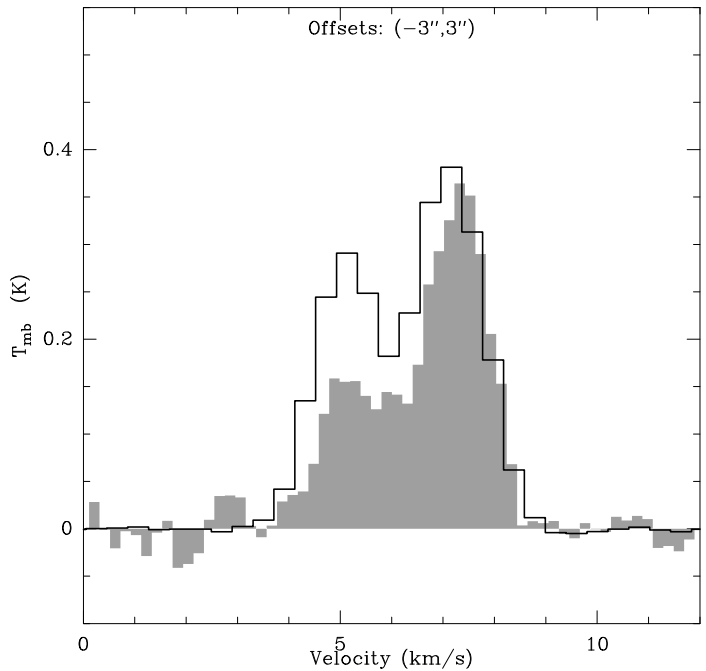}
\caption{Emission spectra of $^{12}$CO(1-0) (grey scale histogram) and \co21\
(thick line) centred at the (-3, 3) offset. 
\co21\ spectra from adjacent positions were convolved to synthetize the \co21\
emission in the 22'' (1-0) Gaussian beam (see text).
Main beam temperatures are adopted.
}
\label{f:sp10-21}
\end{figure}

We now consider the additional 
$^{12}$CO(1-0) spectra (\Fig{sp10-21}) to investigate the 
excitation of the CO gas at the (-3,3) position (at other positions, the
$^{12}$CO(1-0) spectra are of lower S/N due to a smaller integration
time and are less appropriate to measure the excitation ratio). 
For this purpose, $^{12}$CO(1-0) and \co21\ spectra are brought in the
main beam temperature scale; the forward and beam efficiencies are given in
\Tab{CO}.
We next synthetize the \co21\ emission over the (1-0) beam using the 
spectra obtained at 
adjacent positions [(-9,3), (-3,9), (3,3), (-3,-3)] with the appropriate
weighting. The resulting (2-1)/(1-0) integrated intensity ratio 
appears to be around 1.5, a large value compared to that 
($\simeq$ 0.7) commonly observed towards diffuse clouds 
\citep{falgaroneal98,lisztal98,petyal08}. 
Moreover, the (2-1)/(1-0) ratio is significantly different for the two main
components around \vlsr\ $\simeq$ 5 and 7~\kms\ for which we get values of
about 1.8 and 1.3, respectively. 

The above results on the (-3,3) position can be used to set constraints
on the physical conditions 
prevailing in the gas. With the help of a large velocity gradient 
model \citep{hilyblantal07}, we find that acceptable solutions can be 
obtained for $T > 12$~K 
and that above $T = 20$~K, the gas density is well constrained and must lie 
in the range $n = 10^3 - 10^4$\, \cmt\ ($n$ is the ambient H number
density). All solutions obtained correspond to optically thin
emission. 
For instance, at $T = 80$~K, a value close to
the temperature estimated for the dust by \citet{franceal07}, 
we get $n = 2\times10^3$~\cmt\ and 
$N({\rm CO}) = 3.7\times10^{14}$\,\cmd\ 
at \vlsr\ = 5~\kms\ and $n = 1.8\times10^3$\,\cmt\ and $N({\rm CO})
=6.0\times10^{14}$\,\cmd\  
for the \vlsr\ = 7~\kms component. Note that the above CO column densities are
comparable to the value $N({\rm CO}) = 5.7\times10^{14}$\,\cmd\ inferred
either from IUE data by \citet{mclachlanal84} or from FUSE
by \citet{shefferal08} for the gas {\it in front of}
HD~34078.

We end this section by concluding that our CO observations allow us to 
unambiguously answer the question which motived them: HD~34078 is indeed 
closely associated with molecular gas located in its immediate
vicinity (which was previously assigned to a foreground quiescent
cloud by B05). Moreover, the anomalous CO excitation observed, the large
inferred gas density and peculiar velocity field 
very likely result from the interaction between the stellar
wind/radiation and 
the ambient molecular material (in Sect. 6, we discuss in more detail the 
implications of the CO observations).

\section{Variation of CH and CH$^+$ absorption lines}
\label{s:visible}

\subsection{Description of optical observations}

In the visible, we add 26 spectra to the data considered 
in R03. Twelve spectra were taken at OHP, eight at McDonald 
Observatory (hereafter McD) while three spectra were 
obtained at the Boyunsan Optical 
Astronomy Observatory (BOAO, South Korea), two at the Terskol Observatory 
and one at Calar Alto. Altogether, these observations probe the
evolution of CH and 
\CHP\ abundances between 1989 and 
2008, with good sampling since 2000. In particular, our recent 
data cover the period during which the eight FUSE spectra were obtained well.
The date of each observation and spectral resolution are given in \Tab{obs}. 

To check in a direct way the consistency of measurements performed 
at different telescopes, we started in 2003 parallel observations of 
the bright star \zper. The latter has been observed in nearly all 
OHP and McD runs, in addition to HD~34078 (cf \Tab{obs}).
This nearby reddened star [$d = 400$ pc, E(B-V) = 0.33] has a 
small proper motion of 10.2 mas yr$^{-1}$, corresponding to a transverse 
velocity of  4.1 AU yr$^{-1}$, much smaller than the value for 
HD~34078 (22 AU yr$^{-1}$). 
Thus, we expect in principle, much less variation due to structure in the 
foreground IS gas towards \zper\ and absorption lines seen in the spectra 
of that star should be a good indication of the instrumental stability and 
homogeneity of our measurements. Data from the literature  
\citep{allen94,craneal95} do show that $W_{\lambda}$ values for CH, \CHP, CN, 
\ion{Ca}{i}, and \ion{Ca}{ii} are constant within errors for \zper.

OHP observations were done in service mode using the ELODIE spectrograph
\citep{baranneal96}, 
as in R03, except for the four latest runs which were performed with 
SOPHIE, the new spectrograph that now supersedes ELODIE at the 1.93 m 
telescope \citep{bouchyal06}. SOPHIE provides an improved spectral 
resolution ($R$ = 75\,000) 
and sensitivity, as well as an extended wavelength range 
(including CH~$\lambda$3886, 
CH~$\lambda$3890, and \CHPla; as the blue CN lines are close to the blue edge of the 
spectra the S/N is too low for these features to be usable). Since SOPHIE has
been optimized for the detection of extrasolar planets by radial velocity 
measurements, it provides an accurate wavelength scale (better than 0.01
\kms; this scale is relative to the barycentre of the solar system). 
Each observation
consisted in 4 to 8 individual exposures totalling about 1 hr
for HD~34078 and 10$^m$ for \zper; these two targets were generally
observed consecutively or, occasionally, during two sucessive nights. 
Spectra were extracted using the pipeline data reduction software. 
The latter 
were designed specifically for these spectrographs
\citep{baranneal96,bouchyal06} and include all required 
steps (in particular bias and flat field corrections, using appropriate 
exposures). For observations of bright stars such as \zper\ and HD~34078, 
this procedure has been shown to work efficiently, which we checked whenever 
it was possible, e.g. by comparing independent spectra taken at short time 
intervals before merging them, or by comparing $W$ values for those lines 
seen twice, on two consecutive orders.

At McDonald Observatory echelle spectra of HD~34078 were 
obtained with the Harlan J. Smith 2.7 m telescope. The strongest molecular 
features from CN, \CHP, and CH are detected, as are the K line of
\ion{Ca}{ii} and \ion{Ca}{i} $\lambda$4226, although the latter falls next 
to a CCD glitch and thus cannot be reliably extracted. The data were 
reduced in the same fashion as before (cf. R03). A global multi-order fit of the
entire CCD chip was performed for each Th-Ar spectrum, yielding residuals
below 0.001 \AA\ (or 0.08 \kms). Measured radial velocities of (non-variable)
absorption lines towards the comparison star \zper\ show a scatter that is
consistent with the residuals from the wavelength calibration. 
From the Th-Ar data we measured the instrumental resolution of our spectra, 
which turned out to be $R$ = 170\,000. Stellar exposures were 30$^m$ long for 
HD~34078, and typically 5$^m$ for \zper.

A few observations of HD~34078 were also performed at other telescopes. Two
spectra were obtained using the MAESTRO spectrograph, fed by the 2-m 
telescope at the Terskol Observatory (TE) in Northern
Caucasus (the resolution was 120\,000). Three more spectra were obtained using the
fiber-fed echelle spectrograph installed at the 1.8-m telescope
of the Bohyunsan Optical Astronomy Observatory (BOAO) in
South Korea \citep[some description can be found in][]{galazutdinoval05}. 
Modes providing resolutions, $R$ = 30\,000 or 45\,000 were employed (\Tab{obs}
).
Finally, a spectrum was taken in service mode at the 
Calar Alto observatory with the FOCES spectrograph \citep[$R$ = 40\,000;][]{pfeifferal98}.
For these additional runs, \zper\ was not observed; we are 
nevertheless confident that these data are homogeneous with respect to
the other spectra obtained. 
 
\subsection{Equivalent widths and column densities}

\subsubsection{Equivalent width estimates}

\begin{figure}
  \unitlength=1cm
  \begin{picture}(10,9)
    \centerline{\psfig{width=\linewidth,angle=0,figure=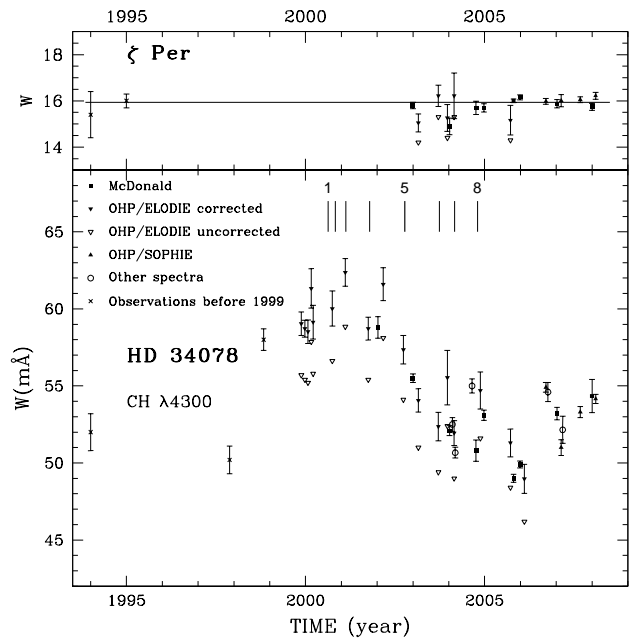}}
  \end{picture}
  \caption{Equivalent widths of \CHl\ measured in HD~34078 
(lower panel) and \zper\ (upper panel) spectra (see \Tab{obs}).
Open triangles correspond to raw OHP/ELODIE values, which appear 
systematically too low (filled triangles correspond to corrected values, 
see text). Tick marks numbered 1 to 8 show the epochs at which the 
FUSE spectra have been taken.}
  \label{f:CHvariation}
\end{figure}

\begin{figure}
  \unitlength=1cm
  \begin{picture}(10,9)
    \centerline{\psfig{width=\linewidth,angle=0,figure=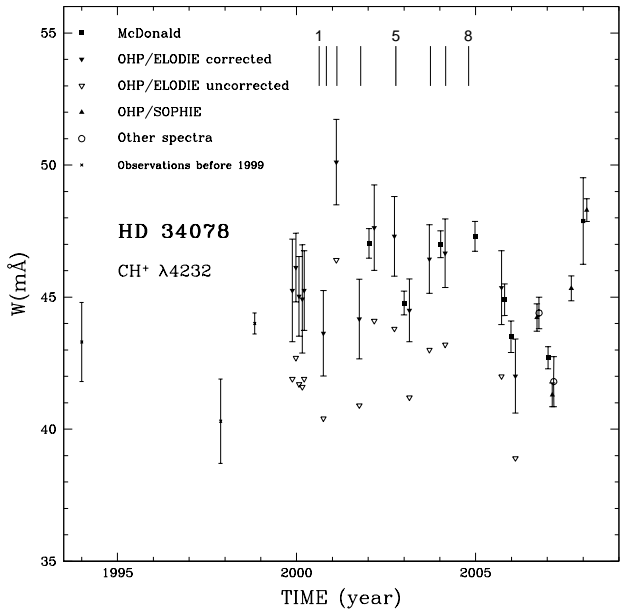}}
	\end{picture}
\caption{Same as in \Fig{CHvariation} but for the equivalent widths of \CHPlb\
in HD~34078 spectra. Raw OHP/ELODIE values (empty triangles) have been 
corrected by + 8\% (filled triangles)}
\label{f:CHPvariation}
\end{figure}

Equivalent width measurements were performed in a similar way as in R03 
for \CHl. Regarding \WCHPlb, the accuracy is limited by the poor 
definition of the continuum due to blending with a stellar line (cf.
Fig. 7 in R03). 
Then, to avoid possible resolution-dependent effects on the estimate of $W$ 
values and to improve the homogeneity and accuracy of our measurements, 
we fitted this stellar line with a Gaussian. Its position 
($\delta\lambda$ = $-0.364$~\AA\ with respect to the \CHP\ absorption) and 
FWHM (0.46~\AA) were fixed while the depth was varied so as
to match the continuum blueward of the \CHP\ absorption as well as
possible (a typical value for 
the depth is 3\% of the continuum). Once the spectra have been
normalised in this way, we measure $W$ and its uncertainty as done for
\CHl. This procedure was applied to all spectra, including those
obtained prior to 2003; for the latter, this results in $W$ values which 
differ slightly from those given in R03.

Some observations of HD~34078 were spread over 2 or 3 consecutive nights
(in particular the recent, high S/N, McD and OHP/SOPHIE spectra), 
which allowed us to check that no significant day-to-day variations 
are present; we then measured $W$ on the co-added spectrum. For the same 
reason, one BOAO and one TE measurement obtained two days apart in March
2004 were combined (thus \Tab{obs} contains 25 entries).

Uncertainties were estimated in a conservative way, including the error due 
to finite pixel-to-pixel S/N ratio and the one in continuum placement; we 
assume as in R03 that the two sources of errors combine quadratically. 
$W$ estimates for HD~34078 and their associated errors are given 
in \Tab{obs} for HD~34078 (\CHl\ and \CHPlb) and \zper\ (\CHl); note that
concerning OHP/ELODIE spectra, values corrected as explained below are 
given for both HD~34078 and \zper.

\subsubsection{Consistency of all measurements}

In \Fig{CHvariation} (upper panel), we show results from the \CHl\ 
\zper\ observations performed between 2003 and 2007, to which we add 
older measurements from \citet{craneal95} and \citet{allen94}. 
These are consistent with a constant value of \WCHl. However, 
careful examination reveals 
a small offset of about -6\% for the OHP/ELODIE values with respect to the 
other measurements. Since the same offset appears to be present in the
HD~34078 \WCHl\ values (lower panel, empty triangles), this effect 
is very likely due to scattered light in the ELODIE spectrograph 
(S. Ilovaisky and P. Prugnel, personal communication). Thus, we applied 
a $+6$\% correction (scattered light from the target does lead to a 
multiplicative correction on $W$ values) to all OHP ELODIE \CHl\ values
(filled triangles), including those presented in R03.

Unfortunately, the \CHPlb\ line towards \zper\  is too faint 
(W $\simeq 2.5$ m\AA) to assess whether OHP/ELODIE measurements of 
this transition are also affected by scattered light. Then, to 
determine the correction for \CHPlb\ (which may be different 
from that for \CHl), we have to rely on the HD~34078 data 
themselves (\Fig{CHPvariation}). By comparing the sets of OHP and 
McD values, we find that an offset of about +8\% needs to
be applied to the OHP/ELODIE values to bring both sets of points 
in good mutual agreement (filled triangles in \Fig{CHPvariation}).
After correcting the OHP/ELODIE $W$ values in this way, it is apparent in 
\Fig{CHvariation} and \Fig{CHPvariation}, that nearly simultaneous 
measurements performed at different telescopes yield consistent values, 
within errors. This is a direct indication that uncertainties are not 
underestimated and that, after correction of the OHP/ELODIE $W$ values,
the whole set of data is homogeneous. 

\begin{figure}
\unitlength=1cm
\begin{picture}(10,9)
\centerline{\psfig{width=\linewidth,angle=0,figure=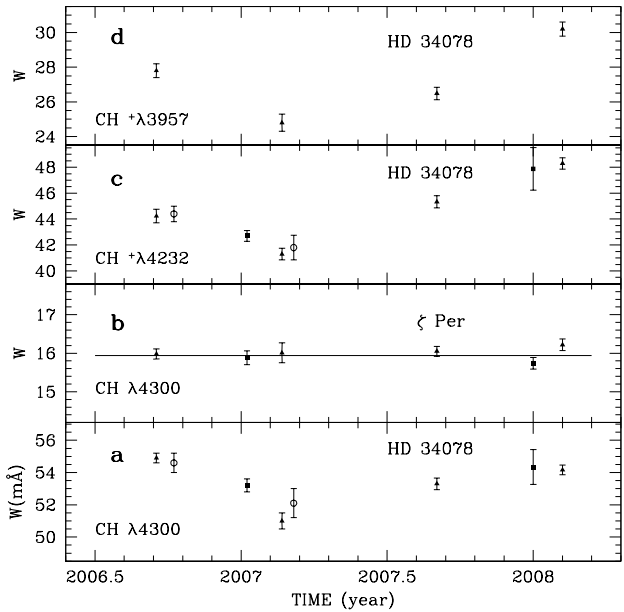}}
\end{picture}
\caption{Equivalent width of \CHl\ (a), \CHPlb\ (c) and \CHPla\ (d) in 
HD~34078 spectra since September 2006 (symbols used have the same meaning as 
in \Fig{CHvariation}). The equivalent width of \CHl\ in \zper\ spectra (b) 
is shown for comparison (the horizontal line corresponds to the weighted 
average of the displayed values). Note the similarity 
between both \CHP\ transitions and the good CH - \CHP\ correlation.}
\label{f:CHCHP_recent}
\end{figure}

The OHP/ELODIE $W$ measurements of \CHPla\ are not accurate enough in 
comparison to those for \CHl\ or \CHPlb\ to be really useful (further, 
they are affected by scattered light in an unknown way). In contrast, 
the recent OHP/SOPHIE spectra provide good S/N values 
for \WCHPla. In \Fig{CHCHP_recent} (panels c and d) we show 
the variation of $W$ versus time for both \CHP\ transitions since 
September 2006; as can be seen, the two sets of measurements are very 
consistent. We also display for the same epochs the behavior of 
\WCHl\ for both HD~34078 (panel a) and \zper\ (b); the latter values
remained constant while the variations seen for CH in HD~34078
are qualitatively similar to those observed for \CHP. The 
whole set of values for \WCHPlb\ and \WCHl\ shows a fairly 
smooth variation, with apparently little or no variations with timescales 
smaller than a few months. 

\Fig{CHvariation}, \Fig{CHPvariation} and \Fig{CHCHP_recent} 
strongly suggest 
that the equivalent widths varied for HD~34078 while \WCHl\ remained 
constant for \zper.
Let us now assess in a quantitative way the statistical significance of 
time changes in the observed $W$ values. 
To this end, we perform a $\chi^2$ test on \WCHl\ values for 
both the \zper\ and HD~34078 sets of measurements, in order to check whether 
the assumption of a constant $W$ can be accepted or rejected. Using the 
whole data set, we find that the \zper\ data are consistent with a
constant value, \WCHl = 15.93 m\AA, while the assumption of a constant 
$W$ value can be rejected at the 5.2 $\sigma$ level for
HD~34078 (the $W$ value which minimizes $\chi^2$ is 52.3 m\AA).
Note that these conclusions are unchanged if the OHP/ELODIE data are removed.
Equivalently, if the distribution of $x=(W - <W>)/\sigma$ values is
compared to a Gaussian with $\sigma_x = 1$, one finds that it is
consistent for \zper\ values and inconsistent for HD~34078 (the $\chi^2$
test described above is just a way to quantify these statements). 
We can therefore conclude that real time variations of \WCHl\ 
occurred during the 2003-2008 period for HD~34078 (for \CHP\ towards
HD~34078, the assumption of a constant $W$ is rejected at the 3.2 $\sigma$ 
level).

\subsubsection{From equivalent widths to column densities}

Since the absorption lines considered here are not completely optically thin,
we cannot infer $N$ values directly from $W$ ones. We will then rely on the
highest resolution data, from McD observations. 
In the latter, the \CHPlb\ profiles are 
fully resolved; one can thus derive the true optical depth and get 
$N$(\CHP) by direct integration.
Regarding CH, the determination of $N$ and of the 
velocity profile is complicated by the 
structure of the ground level, due to $\Lambda$ doubling. As shown by 
\citet{lien84}, neglecting this effect may cause an underestimate 
in $N$(CH) and a broadening of the profile, when the intrinsic 
width is small enough (i.e. comparable or smaller than the splitting of 
the ground level which corresponds to 1.43~\kms). 

CH profiles include at least two components (a strong asymmetric and 
narrow one superimposed onto weak shallow absorption) and fitting with 
Voigt profiles remains somewhat arbitrary (Fig. 5 and 
Fig. 7 in R03). Thus, to get the intrinsic CH pixel optical depth 
profile without any a
priori decomposition and perform the detailed comparison with the 
\CHP\ profile allowed by the quality of the McD data,   
we use a Bayesian inversion procedure as done by \citet{pichonal01}
 in the context of Ly$\alpha$ absorption in QSO spectra. 
Equally populated sublevels are assumed \citep[we checked that the 
relative strength of the CH~$\lambda$3886 and CH~$\lambda$3890 lines in 
the OHP/SOPHIE spectra is consistent with this hypothesis; 
see also][ for other lines of sight]{lien84}. 
We find in the end that taking into account $\Lambda$ doubling induces 
corrections on $N$(CH) which are no larger than 1.3\%.

\begin{figure}
  \unitlength=1cm
  \begin{picture}(10,9)
  \centerline{\psfig{width=\linewidth,angle=0,figure=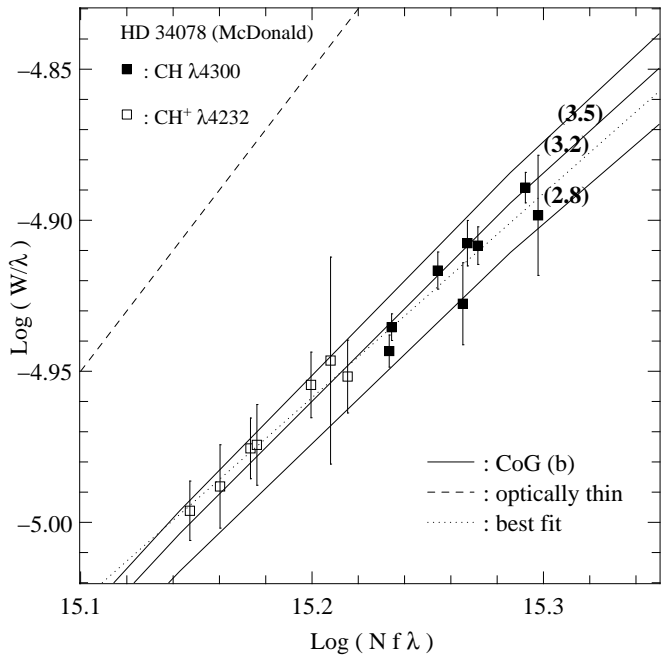}}
  \end{picture}
\caption{Equivalent width versus column density for 
high resolution observations (McDonald). CH (filled squares) 
and \CHP\ (empty squares) absorption 
lines span different ranges in opacity which 
helps to constrain the best linear fit 
(dotted line: $y = -15.232 + 0.676\,x$). A curve of growth with
$b=3.2$ \kms\ also provides a good fit to the data (plain line; the two
 additional lines correspond to $b=2.8$ and $b=3.5$ \kms\ and bracket
 the data points well).
The large $b$ curve (i.e. optically thin limit) is shown (dashed line)}
\label{f:CoG}
\end{figure}

Significant changes in the CH and \CHP\ profiles are seen (see \Sec{CHprofile}) 
but these remain relatively small and further, the optical thickness of 
either \CHl\ or
\CHPlb\ does not exceed 1. One can therefore expect a simple
empirical relation to hold between $W$ and $N$, allowing us to 
infer $N$ values from 
$W$ measurements performed from lower resolution data. 
In \Fig{CoG}, we show 
$W$/$\lambda$ versus $Nf\lambda$ for both \CHl\ and \CHPlb\ from McD spectra. 
As can be seen, both sets 
of points are well fitted by a single straight line (dotted line), as a result 
of the similarity of CH and \CHP\ velocity profiles (cf R03 and
\Sec{CHprofile}). This fit corresponds to the following relations,
\begin{eqnarray}
N({\rm CH}) &=& 2.42 \times 10^{11} W^{1.48} \label{e:fitcog1}\\ 
{\rm and}&&\nonumber\\
N(\mbox{CH$^+$}) &=& 2.29 \times 10^{11} W^{1.48}
\label{e:fitcog2}
\end{eqnarray}

\noindent for \CHl\ and \CHPlb\ respectively, with $W$ in m\AA\ and $N$ in \cmd.
It is noteworthy that although we made no assumption on the shape of
line profiles in the above procedure, the best fit turns out to be close to the
curve of growth for a single gaussian component with $b$ = 3.2 \kms. All
measurements are nicely bracketted by curves of growth with $b$ values between
2.8 and 3.5 \kms. 
We next assume that the $N(W)$ best fit relation holds at all 
epochs and use it to infer $N$ from $W$ for measurements other than 
the McD ones. In \Tab{obs}, we list $N$(CH) and $N$(\CHP) values; for McD
spectra, these are drawn directly from line profiles while for lower
resolution data, $N$ is obtained from $W$ through Eqs.~\ref{e:fitcog1} and \ref{e:fitcog2}. 
The resulting uncertainty in $N$ should in principle involve 
two terms, one related to the errors on $W$ and the other to the scatter about 
the best fit; in practice, the latter appears to be small and we ignore it.

\subsubsection{CH and \CHP\ column density variations}

\begin{figure}
\unitlength=1cm
\begin{picture}(10,9)
\centerline{\psfig{width=\linewidth,angle=0,figure=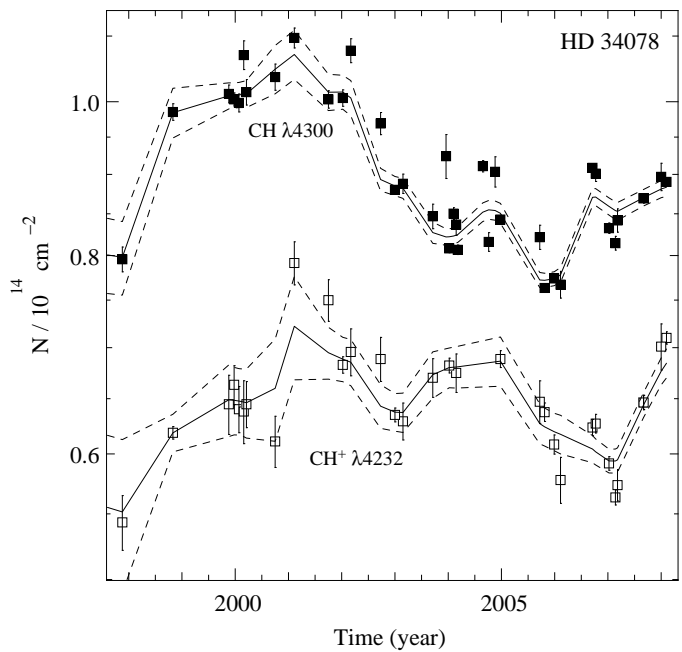}}
\end{picture}
\caption{Evolution of the CH and \CHP\ column densities
with time. The solid thick lines show the long-term evolution obtained after
smoothing using a gaussian window with a FWHM of one year (dashed
lines indicate $\pm 3 \sigma$ bounds on smoothed values).}
\label{f:Nvariation}
\end{figure}

\Fig{Nvariation} shows the variation of $N$(CH) and $N$(\CHP) observed between
1998 and 2007. In R03 we found that \WCHl\ increased from
1990 to 2002; the $+6$\% applied to the OHP/ELODIE data results in a
larger amplitude for this variation (21\% instead of 14\%).
It now appears that $N$(CH) reached a maximum during the 2000-2002 interval,
and has been decreasing since then to reach in 2006 a value similar to those
observed before 1998. To better characterise the long-term variation,
we smoothed the data using a gaussian window with a FWHM of one year.
In the averaging, each $N$ value is weighted 
according to 1/$\sigma^2(N)$ where $\sigma(N)$ is the uncertainty; the 
resulting curve (together with $\pm 3 \sigma$ bounds) is shown in 
\Fig{Nvariation}. The typical timescale for these variations, $\tau_N$, 
defined as the time needed for $N$ to change by 10\% (i.e.  
$\tau_N = [dN/(0.1 N dt)]^{-1}$) is about 1.5--2 yr.

The new data clearly indicate that additional variations, 
with shorter timescales, are also present as was suspected from 
earlier data (R03). 
This is especially clear in the most recent 2006--2008
results: $N$(CH) decreased by $-$10\% between September 2006 and February 2007, 
($\tau_N \simeq$ 0.5 yr) and then increased by 9\% to reach the last
February 2008 value (see \Fig{CHCHP_recent} for a zoom on the recent
$W$ data).

We now consider the \CHP\ measurements shown as empty squares in
\Fig{Nvariation}. The $+8$\% correction applied to the OHP/ELODIE data
somewhat affects the previous conclusion of R03 of a possible decrease 
of $N$(\CHP)
between 1990 and 2002: $N$(\CHP) now looks essentially constant 
over the long term, with no large amplitude variations as those seen for CH.
Beyond 2000, the general pattern is relatively similar for both
species. However, although the 
long-term variations in CH and \CHP\ seem to be loosely correlated, 
there is good correspondence between the short-term ones; this is
especially clear in \Fig{CHCHP_recent} (note in particular the coincidence of
the local minima in February 2007 for both species). Although the variation
patterns are qualitatively similar, the amplitudes are significantly
different. For instance, the increase of $N$(\CHP) between February 2007
and 2008 ($+26$\%) is notably larger than that of $N$(CH) ($+9$\%).

We shall now focus on the high resolution spectra of the \CHl\ and \CHPlb\
absorption lines in order to investigate whether the observed long and 
short-term variations in $W$ are accompanied by detectable changes in line 
profiles. Profile variations might also be present which are not necessarily 
reflected in appreciable $W$ changes.

\subsection{High resolution CH and \CHP\ line profiles and their variations}
\label{s:CHprofile}

The high resolution profiles obtained at McDonald Observatory have all been 
corrected for the Doppler shift due to the Earth velocity and have been 
brought in the LSR system. Yet, after correction, the 
wavelength of the \zper\ CN and CH lines still show slight fluctuations 
in position (of at most a few m\AA) from one epoch to another (given the
excellent S/N and the sharp line profiles, misalignments by only a few
m\AA\ are sufficient to induce significant profile differences). These 
fluctuations are identical for both CN and CH lines to within 1 m\AA\ 
and show a good correlation with those of the HD~34078 lines. 
They must then be due to some inaccuracy in the LSR correction from one 
epoch to another. To improve the accuracy in the alignment of the 
HD~34078 absorption lines, we assume that \zper\ lines have been stable 
in position (this is confirmed by SOPHIE spectra whose wavelength scale 
is accurately defined) and infer the value of the relative shifts for 
each epoch. These are used to slightly adjust the position of 
HD~34078 lines.

\begin{figure}
\unitlength=1cm
\begin{picture}(10,12)
\put(0,0){\psfig{width=17cm,angle=-90,figure=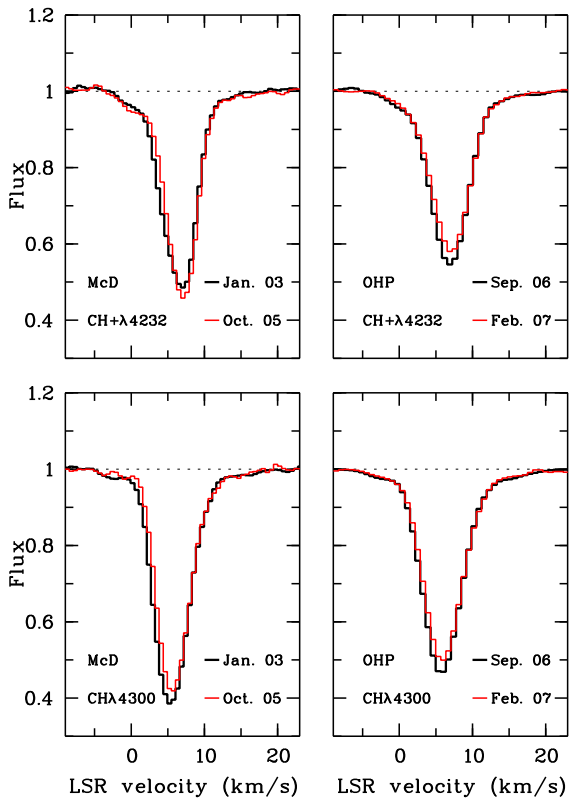}}

\end{picture}
\caption{Comparison of McD \CHl\ (lower left) and \CHPlb\ (upper
 left) profiles observed in January 2003 and October 2005. 
During this 2.81 yr time interval $W$ has varied 
by $-$11.7\% and +0.3\% respectively. Same for OHP/SOPHIE 
\CHl\ (lower right) and \CHPlb\ (upper right) profiles obtained 
 in September 2006 and February 2007 ($W$ variation
of -7.1\% and - 6.6 \% over 0.43 yr). Note the good stability of the red
 side of the profiles.}
\label{f:varprof_mcdohp}
\end{figure}

The CH and \CHP\ line profiles are quite similar (\Fig{varprof_mcdohp}): both 
include a narrow component with a full width at half maximum of about 5.5 \kms\ 
(corresponding 
to a $b$ parameter of 3.3  \kms) and shallow absorption extending 
from $-$4 up to 17 \kms. 
The CH and \CHP\ narrow components cover the same velocity range 
($\simeq$ 2 - 11 \kms) 
but their shapes are significantly different: CH displays a steeper blue
edge, while the opposite holds for \CHP. This results in a shift of
about $+1.6$~\kms\ for the \CHP\ line centre with respect to that of CH. 
It is noteworthy that the velocity range covered by the narrow CH or \CHP\
absorption coincides quite well with the range over which CO emission 
is observed.

Since we are likely probing molecular gas closely associated with
HD~34078 (i.e. with peculiar physical conditions), one may wonder
whether the CH and \CHP\ profiles show significant differences with those seen on other
lines of sight. We simply note that the presence of a weak broad component is rare 
\citep{craneal95,crawford95} but such shallow absorption would 
be difficult to detect in most spectra; one of very low amplitude 
($\simeq$ 1\% instead of $\simeq$ 5\% in our spectra) is however present
in the very high S/N spectrum of $\zeta$ Oph presented by \citet{craneal91}. 

\begin{figure}
\unitlength=1cm
\begin{picture}(10,10)
\put(0,0){\psfig{width=17cm,angle=-90,figure=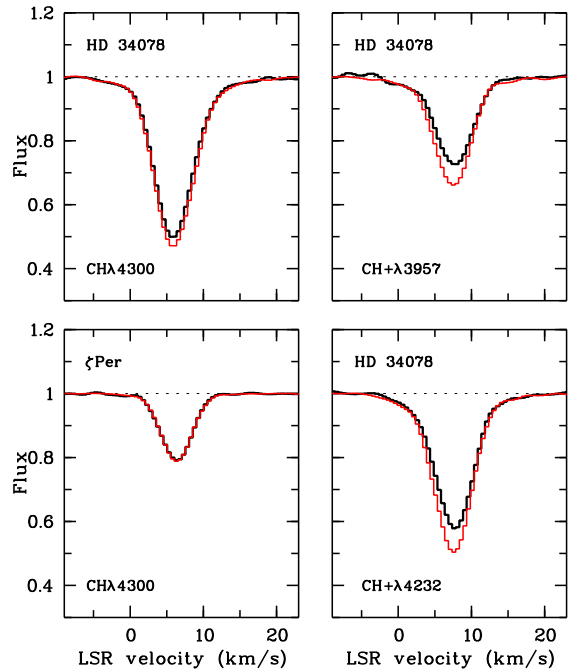}}
\end{picture}
\caption{Comparison of the HD~34078 OHP spectra obtained in February
 2007 (black) and February 2008 (red) for the \CHl\ (upper left),
 \CHPlb\ (lower right) and \CHPla\ (upper right) lines. Note the good
 consistency between the two \CHP\ transitions. During this 1.0 yr
 interval, the $\zeta$ Per \CHl\ profile remained stable (lower left).}
\label{f:varprof_ohp}
\end{figure}

By comparing successive spectra, we find that fluctuations in $W$ are
most often associated 
with changes in the blue and central part of the profiles, the red side
suffering very little
variation. This is apparent in \Fig{varprof_mcdohp} (left panels)
in which we display 
CH McD profiles for the January 2003 and October 2005 epochs. During 
this 2.8 yr time interval \WCHl\ has decreased 
by 12\% and the FWHM of the narrow component decreased by 5\%. 
\WCHPlb\ remained nearly constant over the same period but
significant profiles changes are nevertheless clearly present; the line
is slightly deeper, which compensates for a decrease in FWHM 
comparable to that seen for CH. Another example is shown in
\Fig{varprof_mcdohp} (right panels), where we compare OHP/SOPHIE CH and \CHP\ profiles 
taken over a much shorter interval (0.43 yr from September 2006 to
February 2007). 
In this case,\WCHl\ and \WCHPlb\ have decreased by a comparable
amount ($\simeq$ 7\%) while both profiles became slightly narrower. 
Thus correlated $W$ variations appear to be associated with similar
profile changes. Since 2003, $W$ values show no systematic trend but
rather erratic fluctuations; the same is true for profiles and
generally, for two epochs with comparable $W$ values, the profiles are 
quite similar (the \CHPlb\ January 2003 and October 2005 profiles being an
exception). In \Fig{varprof_ohp} we display the recent February 2007 and 
2008 CH and \CHP\ profiles; the latter show a marked increase in $W$ (cf
\Fig{CHCHP_recent}), corresponding to $+26$\% for $N$(\CHP)~! (Note that
during the same interval, the \CHl\ \zper\ profile remained stable.) We
do not see appreciable variations of the broad shallow
component, but given its weakness, the significance of this result is
limited.

Unfortunately, the S/N ratio for the CN line profiles of HD~34078 is not 
sufficient to allow a search for variations with a sensitivity
comparable to the one attained for CH and \CHP. 
Further the strongest CN line is clearly affected by variations of blended 
stellar (\ion{C}{iv}) absorption. Thus we shall not discuss variations of CN features.

\section{Variations in the \H2\ column density}
\label{s:FUSE}

\cite{boisseal05} analyzed the first 5 FUSE spectra taken since 2000 
and detected no variation for the \H2\  column density with an upper limit of
5\% at a 3$\sigma$ confidence level. Three additional spectra were 
obtained by FUSE in September 2003, February 2004 
and October 2004 which allow us to follow the time behavior of \H2\
absorption during nearly five years since January 2000 and then probe the spatial
distribution of the gas over scales up to 104~AU. 

\begin{figure}
  \unitlength=1cm
  \begin{picture}(10,7)
    \put(-1,0){\psfig{width=10cm,angle=-90,figure=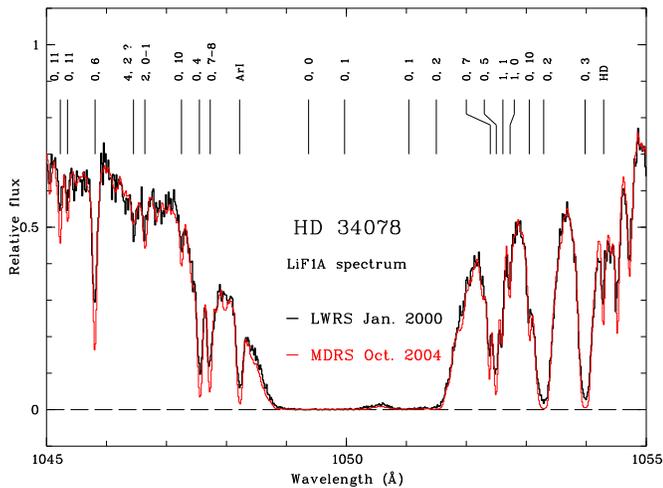}}
  \end{picture}
  \caption{Comparison of the FUSE spectra obtained in January 2000
 (LWRS, black line) and October 2004 (MDRS, red line) in the region around
 1050 \AA\ ($v$ and $J$ values for the lower level of each \H2\ transition
 are indicated; the HD line is from $J=0$). 
In the MDRS spectrum, narrow lines tend to appear
 deeper and the damped (0,0) and (0,1) features slightly broader.}
  \label{f:sp1-8}
\end{figure}

\subsection{Contamination of HD~34078 spectra by nebular emission}

Comparing the October 2004 (8$^{\rm th}$) spectrum to the
previous ones, we find that it shows noticeable differences
(\Fig{sp1-8}): narrow lines are deeper and damped \H2\ lines are slightly broader
[as if N(\H2) had increased]. This brought to our attention a potential
difficulty that had not been considered in B05: contamination of the
HD~34078 spectrum by diffuse emission from the IC~405 nebula.
Indeed, the 8$^{\rm th}$ spectrum was obtained with the MDRS 
aperture while all others were taken using the larger LWRS aperture.
Given the difference in size (MDRS: 4 $\times$ 20 arcsec$^2$; LWRS:
30 $\times$ 30 arcsec$^2$) and the intense diffuse emission detected
close to the HD~34078 line of sight by \citet{franceal04},
the peculiarities of the 8$^{\rm th}$ spectrum might just be due to a
lower level of contamination of the spectrum by
diffuse emission. 
In Appendix A, we estimate the nebular contribution to LWRS spectra and  
conclude that it can explain the difference between FUSE 
spectra 1 to 7 and the 8$^{th}$ MDRS one.Thus only spectra 1
to 7 will be considered below in our search for variations in \H2\ lines.

The importance of diffuse emission contamination in LWRS spectra also
implies some limitation in our search for variations: the aperture may
not be located exactly at the same position on the sky at all epochs
resulting in a slightly variable contribution from diffuse emission if gradients
are present. Note that since the diffuse to stellar flux
ratio decreases with wavelength, \H2\ systems at longer
wavelengths are best suited to minimize the contamination by diffuse emission.
Regarding the study of the gas properties towards HD34078, the
8$^{\rm th}$ spectrum is clearly to be preferred for two reasons: i) it 
should be
much less affected by diffuse emission and ii) the S/N ratio is significantly
higher than for previous spectra due to an integration time (22\,500\,s) about
four times longer than at epochs 1 to 7 ($\simeq$ 6\,000\,s). 
A redetermination of the gas properties based
on the 8$^{\rm th}$ spectrum (\H2\ excitation diagram in particular) 
will be presented
elsewhere; we simply note here that the detection of absorption lines 
from all excited \H2\ levels quoted in B05 is confirmed.

\subsection{Variations in N(\H2, J=0)}
As in B05, we perform a direct comparison of the LWRS spectra,
after relative flux intercalibration and adjusment of the
wavelength scale. This procedure is applied independently
to three portions of the spectra located at about 
1050, 1063 and 1078 \AA, corresponding to
the (4-0), (3-0), (2-0) \H2\ Lyman bands respectively.
Each of these broad features is a blend of four \H2\ lines arising
from the $J=0, 1 $ and 2 levels. A good relative flux
calibration is easily obtained (as for spectra 1 to 5),
indicating that the shape of the stellar spectrum does not
vary \citep[known stellar lines for such O9 stars are indeed
weak and rare in these regions:][]{pellerinal02}.
Using narrow high $J$ \H2\ lines adjacent to the broad
\H2\ absorptions of interest, we get an accuracy in
the wavelength alignment of about 0.01~\AA\ for the
1050, 1063 and 1078 \AA\ absorption systems.

\begin{figure}
\unitlength=1cm
\begin{picture}(10,6.5)
\centerline{\psfig{width=\linewidth,angle=-90,figure=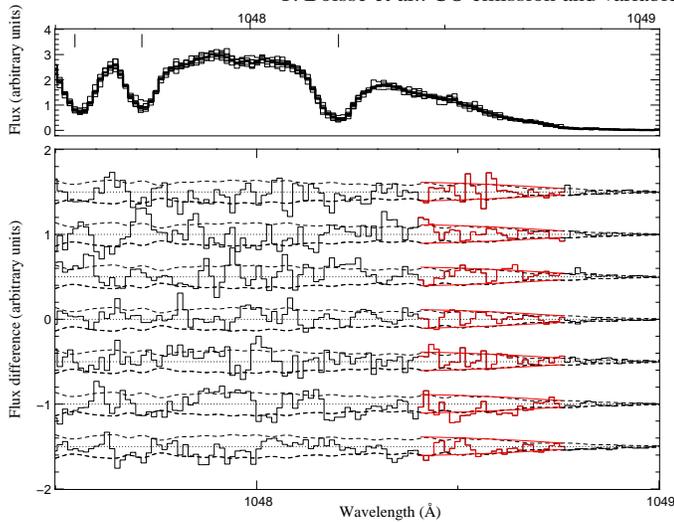}}
\end{picture}
\caption{FUSE LWRS HD~34078 spectra near the 1050 \AA\ broad (4-0) \H2\ 
Lyman band.
{\em Upper panel} : all epochs, after flux intercalibration and
alignment (see text). The mean spectrum is shown (thick line); tick
 marks indicate the features shown in \Fig{sp1-8} (two are from \H2, the
 third is from Ar~I).
{\em Lower panel} : difference between the spectrum for each epoch
and the mean spectrum
(epochs 1 to 7, shifted from top to bottom for clarity; same scale as
in the upper panel).
The dashed line shows the 3$\sigma$ variance among the different
epochs. The region used in the analysis is shown in red.}
\label{f:fuse1050}
\end{figure}

We now focus on the blue edge of each broad \H2\ system
which presents good sensitivity to changes in $N$(\H2, $J=0$). A zoom
of this region for the 1050 \AA\ system is shown
in \Fig{fuse1050} (upper panel). All spectra, corrected in flux and
wavelength as described above, are superimposed. They are all
similar and an average spectrum can therefore be computed
(thick line). The difference ($\Delta I_i$) between one individual
spectrum, $i $, and the mean is displayed in the lower panel
for each epoch ($i=1, ..., 7$ from top to bottom).
The 3$\sigma$ dispersion on
$\Delta I_i$ among the seven epochs is indicated as a function
of wavelength (dashed lines). Away from the $J=0$ line, the
$\Delta I$ profile is consistent with no variation. In the region
close to the $J=0$ line (displayed in red) where variations in
$N(J=0)$ would induce changes in the profiles, spectra 1 to 7
are also consistent with the mean spectrum. 
Similarly, B05 have adjusted the first spectra
using $\log N(J=0)=20.52$ and concluded that the variation
among the five first spectra was lower than 5\%. Indeed, 
an increase (decrease) of this amplitude roughly corresponds to
a difference that follows the lower (upper) 3$\sigma$ profile 
in Fig. 10. We conclude from our analysis, that $N$(\H2) changed 
by less than 5\% at the 3$\sigma$ level between January 2000 and 
February 2004 while $N$(CH) has undergone variations as large as 
20\% over the same time interval (cf Fig. 6).

\section{Observations summary}
\label{s:summary}

\subsection{The stability of $\zeta$ Per CH absorption}
As mentioned above, \zper\ was observed primarily to test the
homogeneity of our
measurements. After correction of the small offset found in the
OHP/ELODIE
data, the whole set of $W$ values appears remarkably consistent. The
latter, and
the most recent (2006 - 2008) data in particular, lead us to a
strong twofold conclusion: i) \WCHl\ remains
constant towards \zper\ and ii) $N$(CH) and $N$(\CHP) do vary 
towards HD~34078. Indeed, if variations of instrumental
origin were responsible for changes in HD~34078 values, one should 
invoke a very unlikely ``conspiracy'' to explain the stability of \zper\ lines.
The fact that the same behavior is observed for distinct \CHP\ lines
is a strong additional proof of the reality of HD~34078's variations.

\zper\ was observed during 5 years and over this time interval, the
drift of the line of sight
through the foreground cloud has been significant. The distance
to the cloud is thought to be 350pc \citep{hiltonal95} and thus the 
drift of the line of sight amounts to 17.8 AU. 
The constancy of $N$(CH) then shows that over this 
scale and below, there is no marked structure in the cloud. The 
3$\sigma$ upper limit on relative variations of $N$(CH) is about 6\%. We
derive this value simply from the raw average and rms scatter of McD 
and OHP/SOPHIE measurements, assuming the \CHl\ line is optically thin; 
a more detailed analysis of the \zper\ data, including
CN and \CHP\ lines, will be presented elsewhere.

\subsection{Main observational results on HD~34078}

From multiwavelength observations of the gas towards HD~34078 and in
its close environment, we get the following results:

\begin{itemize}
\item{the $^{12}$CO(2-1) emission map of the HD~34078 field
     shows a pronounced peak coincident with the star's position, clearly indicating
that molecular gas seen in absorption is closely associated with HD~34078.
The extent and morphology of the CO emission correlates well with the
     24~$\mu$m dust emission arc of \citet{franceal07}. Presumably as a
     result of the interaction between HD~34078 and the ambient cloud, the
CO(2-1)/CO(1-0) ratio is anomalously large, pointing towards dense ($10^3 -
     10^4$~\cmt) and warm ($T > 12K$) emitting gas and further, a 
  remarkable kinematical
     pattern with doubled-peaked profiles is observed,}
\end{itemize}

\begin{itemize}
\item{we confirm the reality of rapid, large amplitude (typically
     10\%~yr$^{-1}$) and correlated variations of $N$(CH) and $N$(\CHP)
towards HD~34078.
The velocity ranges covered by CH and \CHP\ narrow absorption coincide
well with that of CO emission. Variations in CH and \CHP\ line
profiles are unambiguously detected; these occur mainly in the blue
part of the narrow absorption. A broad shallow and relatively stable
     component is seen for both CH and \CHP\ in the interval [-4, 17~\kms],}
\end{itemize}

\begin{itemize}
\item{comparison of LWRS and MDRS FUSE spectra reveals that the 7 LWRS
     spectra available are significantly contaminated by diffuse light
     from IC~405. The absence of variations in the LWRS profiles of \H2\ $J=0$ lines
yields a 3$\sigma$ upper limit of 5\% on $N$ values, extending the
     result of B05 over nearly four years (or 90~AU).}
\end{itemize}

\section{Discussion}
\label{s:discussion}

Given the marked contrast between the stability of \zper\ CH lines and
the rapid, large amplitude variations seen for CH and \CHP\ towards 
HD~34078, we shall 
assume in the following discussion that these variations can be
attributed entirely 
to phenomena associated with the star/cloud interaction and not to small
scale structure in cold gas. 

\subsection{Towards a coherent picture of the close environment of
HD34078}

From the broad set of observations available, a coherent scenario
emerges which can be summarized as follows. HD~34078 recently 
encountered a molecular cloud, as originally suggested by 
\citet{herbig58}. The stellar wind impacts the ambient material resulting
in a shell of compressed, highly excited gas. Modelling of both
the \H2\ excitation (B05) and CO emission (Sect. 2.2) provides 
consistent estimates for the density in the shell, $n \simeq 10^4\,
\cmt$. Dust grains located in this region are 
directly exposed to the intense UV flux from the O star and strongly emit at
infrared wavelengths. Thus, both the mid-IR emission detected by
\citet{franceal07} and the CO emission probably delineate the part of
the shell seen edge-on by the observer, accounting for the good
correlation between the 
{\it Spitzer} 8 or 24 $\mu$m arc and the CO map (Fig. 1).

B05 favored a two-cloud model on the basis of the dichotomy in the 
physical conditions derived for the highly excited \H2\
on the one hand and for all other molecular absorption on the other
hand. They discarded a single cloud model (although it would have
naturally explained the similar velocity for all absorption lines and the 
presence of preexisting diffuse \H2\ near HD~34078 ...) 
because they implicitly assumed that molecules 
should be photodissociated at the very small distance (a few 0.01\,pc) 
implied by
the modelling of the \H2\ excitation. In fact, this argument is valid 
only in a model that is stationary regarding the formation/destruction of 
molecules. Given the large space velocity ($V_s$) of
HD~34078, its arrival is so recent that probably, no such steady-state
equilibrium could be established. Rather, as the O star is approaching the
cloud, a photoionisation and photodissociation front develops, moving at
velocities of the order of a few \kms\ only \citep{bertoldial96}, 
i.e. well below the star velocity. The distance between HD~34078 and 
these fronts then gradually decreases and the velocity of the latter
gets higher (the front velocities increase with stellar flux), 
up to a point where the star and front velocities equate. 

\begin{figure}
\unitlength=1cm
\begin{picture}(10,4)
\put(0,0){\psfig{width=8cm,angle=0,figure=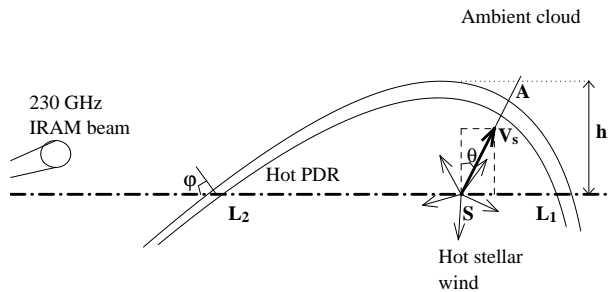}}
\end{picture}
\caption{Cartoon showing some geometrical parameters associated with the 
bow shock around HD~34078 ($h$ is the apparent standoff distance; $h
 \simeq 0.04$ pc). 
The velocity vector $V_{s}$ shown as the
bold arrow is inclined by $\theta = 30\degr$ with respect to the plane
 of the sky. ``S'' marks the star position while the dot-dashed
 horizontal line represents the line of sight to the star (the observer
 being on the left). The double curve is the projection of the proposed 
bow shock. Optical and UV spectra of HD~34078 probe the bow shock/hot
 PDR at $L_2$ (with an
 inclination angle $\varphi$; we estimate $SL_2 \simeq 0.1$ pc) and the 
foreground ambient cloud while IRAM 230 GHz observations involve 
all material enclosed into the beam.}
\label{f:cartoon}
\end{figure}

Moreover, the stellar wind has a strong mechanical impact on the 
surrounding gas and the latter is gradually set into motion as a result
of momentum flux. When the star is close enough to the cloud, a 
stationary bow shock is established at a position nearly coincident with
that of the photodissociation front. At this stage, the distance ($R_0 =
AS$, see \Fig{cartoon}) 
between the star and the apex of the shock is well defined and remains 
constant as far as the density of the ambient cloud and the wind
properties (mass loss rate and terminal velocity) do not change. 
In this picture (\Fig{cartoon}), whether the dynamical steady-state 
regime is 
established or not, the molecular material located beyond the 
front/shock surface should be little affected by the presence of the 
closeby star, thereby accounting for the characteristics of the molecular 
components that B05 assigned to the ``translucent'' component. 

We conclude that a model involving a single cloud located very close to
HD~34078 may, in fact, be consistent with all observations. A key issue
remains open however: at which stage of the
star-cloud interaction have we captured HD~34078 and its close environment ? 
A stationary dynamical regime is expected to establish ultimately and in the
thin-shell limit, detailed models describing the 
geometry and velocity structure of steady-state bow shocks are available
\citep[e.g.][]{vanburenal90,mclowal91,wilkin96}. We thus performed 
a detailed comparison between the predictions of these models and
observations (Appendix B). From this analysis, we 
conclude that the IR/CO arc does not display the properties expected for
a stationary bow shock and that we are possibly observing a ``nascent
bow shock'', i.e. the wind/cloud interaction at an early evolutionary
stage, well before the formation of a steady-state flow around the
star. We also find that the geometrical constraints provided by the IR
data, $h \simeq R_0 \simeq 0.04$ pc, are roughly consistent with the radiation
field implied by the modelling of \H2\ excitation (B05).

\begin{figure}
\unitlength=1cm
\begin{picture}(10,5.5)
\put(0,0){\psfig{width=8cm,angle=0,figure=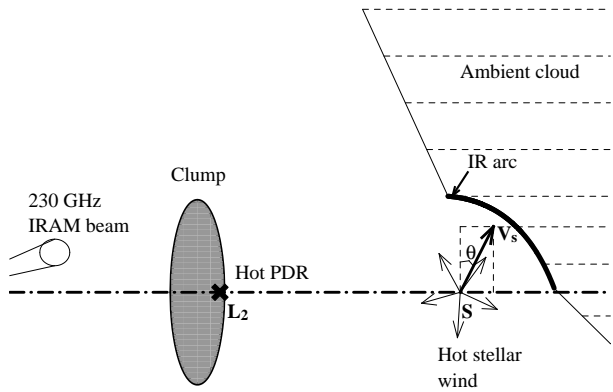}}
\end{picture}
\caption{Alternative scenario to the one presented in
 \Fig{cartoon}. Here, HD~34078's wind slightly modifies
 the shape of the ambient cloud boundary, giving rise to the IR arc.
The clump reddens the star but not the surrounding nebular emission;
 further, gas located on its cold side (facing the observer) induces
 narrow absorption in the \co21\ emission profiles as well
 as some visible and UV absorption lines. Excited \H2\ lines and a significant fraction
 of CH and \CHP\ absorption originates from the PDR located on the
 hot side of the clump, intersected by the line of sight at $L_2$.}
\label{f:clump}
\end{figure}

A variant of the above scenario is suggested by \citet{franceal04}, who 
proposed that differential extinction is present
between HD~34078 and the surrounding diffuse emission, 
in order to explain the increase of the diffuse to stellar ratio
at far-UV wavelengths. 
More specifically, \citet{franceal04} propose that a small clump lies in front of
HD~34078 (\Fig{clump}). 
With an extent no larger that about $20~\arcsec$, the latter could
induce the observed HD~34078 extinction without affecting the
surrounding nebular emission. 

Such a picture is attractive in our context
because $20~\arcsec$ is approximately the size of the area over which
a ``dip'' is seen in the CO profiles, suggesting that the
double-peaked line shapes might be due to narrow absorption rather than
to velocity structure in the emitting gas. In this scenario,
CO emission could originate from the ambient cloud background to HD~34078
 - especially its outer boundary, compressed by the wind - accounting for
the widespread emission seen over most of the field (\Fig{clump}). 
Additional emission 
could come from the clump itself, in particular from the region located
immediately beyond the hot PDR facing the star, explaining the enhanced
emission close to HD~34078 and the high excitation of the emitting material.
Gas located on the cold side of the clump facing the observer,
should be little affected by the 
interaction, as in the bow-shock scenario. This region corresponds to
the translucent component in B05's model; with a small CO excitation 
temperature and low velocity dispersion, this gas could induce narrow 
absorption in the background CO emission. The close similarity of the CO dip 
velocity at the star position, $V \simeq 6.0\,\kms$ (this value is
stable to within 0.2 \kms\ among spectra displaying the dip), and the
velocity of maximum narrow CH and \CHP\ absorption (see \Fig{varprof_mcdohp})
is consistent with absorption being responsible for the dip in CO 
profiles. The clump should be located at a distance from the star of about
0.1 pc (comparable to that of point $L_2$ in the bow-shock picture)
so that the hot PDR facing the star is exposed to a radiation field
large enough to account for the \H2\ excitation (Appendix B, B05). In
fact, the bow-shock and clump scenario are more or less equivalent, both
of them involving the presence along the line of sight of a shell of 
dense gas illuminated by intense UV radiation; the main
difference is that in the clump picture, the ambient material is not
distributed in a continuous manner around HD~34078.

\subsection{CH and \CHP\ abundance and time variations}
In the picture described above (\Fig{cartoon}), the dense shell at the wind/cloud
interface is nearly static in the ambient cloud's frame and as
time elapses, the point ($L_2$) at which the line of sight intersects 
the shell drifts over the latter towards the apex ($A$) at a velocity 
$V_{\mbox{t}}/\cos\varphi$, where $V_{\mbox{t}}$ is the transverse
velocity and $\varphi$ is the inclination of the 
shell at $L_2$. Similarly, in the clump scenario (\Fig{clump}), the line
of sight
drifts over the wind/clump interface. The rapid CH and \CHP\ 
variations observed imply that a significant fraction of these species
is enclosed in a very localised region (with a size of about 10 AU,
corresponding to a time interval of 6 months) and in our context, it is natural to assume 
that time variations arise from structure over the shell of compressed
gas intersected by the line of sight. Further, as
already outlined by B05, the CH/\H2\ ratio towards HD~34078 is
anomalously large: for $N$(CH) $\simeq 10^{14}\, \cmd$ (the largest values 
reached, in 2000 - 2003) we get a ratio of $1.6 \times 10^{-7}$, 
which appears to be 3.7 times larger than the value inferred from the 
best fit given by \citet{shefferal08} \citep[a similar result is obtained 
by comparing HD~34078 values to those compiled by][]{weltyal06}. 
Since the CH - \H2\ correlation is quite good, such a
deviation is very significant and indeed, in the Fig. 8 presented by
\citet{shefferal08}, HD~34078 is clearly an outlier (note that
surprisingly, HD~37903, the other star showing highly excited \H2, has
an anomalously {\it low} CH/\H2\ ratio !). The \CHP/\H2\ ratio displays
much more scatter among all sightlines, but nevertheless, $N$(\CHP) 
towards HD~34078 is among the largest values for $N$(\H2) $\simeq 6 \times
10^{20}\, \cmd$ \citep[Fig. 10 in][]{shefferal08}. In the end, the CH/\CHP\ 
ratio in HD~34078 is well within the range
observed towards other stars \citep[cf Fig. 10 in][]{weltyal06}. We
thus apparently have a comparable relative excess of both CH and \CHP, 
and again, it is natural
to assume that the overproduction of these species occurs in the
compressed shell. To induce such a large deviation in the CH/\H2\ ratio,
the overproduction of CH must be quite large since molecular
gas located beyond the shell probably displays a more standard ratio. 
Similarly, the spatial
distribution of CH and \CHP\ over the shell must be very inhomogeneous
at scales of about 10 AU to induce the observed variations. Since the 
CH and \CHP\ variations are strongly correlated, a single mechanism must 
be at work to explain the overproduction of both species. 

Clearly, the CH and \CHP\ time variations cannot be attributed 
to pure density structure. Indeed, no corresponding changes have been
seen for $N$(\H2) and further, explaining CH or \CHP\ variations by 
more or less spherical clumps with a size of about 10 AU would require 
very high volumic densities, $n$(\H2) $\simeq 10^{5-6}$~\cmt, as 
argued by R03. In such
gas, \CHP\ would be rapidly destroyed by reactions with \H2. Then, the
structure is likely to be more {\it chemical} in nature.
 
The production of \CHP, which is not expected at 
thermal equilibrium, together with the correlated variations of CH and
\CHP, suggest that both molecules are mainly formed in the dense shell, 
through a MHD shock where the drift velocity between ions and neutrals trigger the formation
of both \CHP\ and CH \citep{pineaual86,floweral98}. 
\CHP\ could also be overproduced at the interface between the ambient cloud and
warmer gas, as suggested by \citet{duleyal92}, \citet{crawford95} and more
recently by \citet{lesaffreal07}. 
In the specific conditions prevailing around HD~34078, one may 
imagine that Rayleigh-Taylor instabilities develop efficiently (e.g. as
a result of fluctuations in the wind properties or in the ambient
cloud), leading to the formation of a turbulent mixing layer with
pronounced small-scale structure. HD~34078 being variable
\citep{marchenkoal98}, stellar flux variations can also trigger or
amplify the formation of small-scale structure at the interface and in
the PDR. In regions hot enough to form significant amounts 
of \CHP\ via the endoenergetic C$^+$ + \H2\ reaction
(whatever the heating mechanism, shocks or turbulent dissipation), 
small spatial temperature fluctuations will result in appreciable
variations in the \CHP\ formation rate and then in the relative
abundance of \CHP\ and CH (recall that CH is easily formed from
\CHP\ once the latter species is present). 
One can indeed estimate that since the \CHP\ formation rate
scales as $\exp(-4640/T)$, a local variation as small as 
22~K around $T \simeq 1000$~K is sufficient to induce a 
fluctuation of 10\% in the local abundance.
The broad, shallow CH and \CHP\ absorption components
(\Fig{varprof_mcdohp}) could 
just be a signature of the highly turbulent velocity field at the 
interface (note the excellent agreement in 
the velocity intervals [-4 to 17 \kms] covered by the CH and \CHP\ broad
components, indicating that the species responsible for that component
are cospatial). 

CH and \CHP\ line profiles vary mainly on their blue
side, at \vlsr\ $\simeq 4 - 5$~\kms (\Fig{varprof_mcdohp}), suggesting this velocity for
the gas lying at the interface near $L_2$. This material is therefore
blue-shifted with respect to the ambient cloud (at \vlsr\ $\simeq
6.5$~\kms); this is consistent with the stellar wind pushing
ambient molecular gas towards the observer. The higher excitation
derived for the CO component at \vlsr\ $\simeq 5$~\kms\ also nicely fits
this view. Thus, the scenario described above might, at the same
time, explain in a coherent way the large abundances of CH and \CHP\ 
as well as their time variations.

The amounts of excited \H2\ ($J=3$ to 5) and \CHP\ are known to
correlate \citep{lambertal86} and is taken as an indication that
the same mechanism is responsible for these species, the production of
which requires an input of additional energy of yet unknown
nature. Towards HD~34078, a marked excess of both excited \H2\ and
\CHP\ is observed, in agreement with the correlation seen among more
standard lines of sight. We simply note that in our scenario, the large
amount of excited \H2\ and \CHP\ is a direct consequence of the proximity of
HD~34078, through its wind and UV flux.

\subsection{Small scale structure in quiescent \H2\ gas}
In the picture that we propose, \H2\ gas located beyond the
photodissociation front and shocked region is essentially unaffected by
the presence of the star (dust grains should be somewhat warmer due to the
proximity of the star, but at densities of a few $10^2$ \cmt, this has
little impact on the gas temperature). This part of the cloud gives by far the
dominant contribution to the \H2\ column density in the $J=0$ level 
(in B05's model, the hot PDR represents less than 1\% of the total $N$
value); it should also contain a significant fraction of the CO
responsible for the UV absorption, about 1/4 of CH molecules (cf above) 
and most of the CN. Thus, the lack of variation in $N$(\H2, $J=0$) at a level better than
5\% implies that no marked small-scale {\it density} structure is
present within the fraction of the ambient cloud probed by the drift
of the line of sight between 2000 and 2004. If this region is
representative of quiescent diffuse molecular material in general, the
structure seen elsewhere for other tracers like H$_2$CO, HCO$^+$, and OH would be
mainly ``chemical'' structure, possibly reflecting the specific
formation/destruction processes relevant to these species. 

We further note that the lack of structure in quiescent \H2\ gas is
consistent with the stability of CH absorption lines towards \zper, provided the CH/\H2\
abundance ratio is uniform at scales of about 10~AU within this quiescent cloud.

\section{Conclusions and prospects}
$\bullet$ By mapping \co21\ emission around HD~34078, we have unambiguously
shown that the molecular material seen in the foreground is closely
associated with the star, supporting the
suggestion by \citet{herbig58} that HD~34078 currently encounters a molecular
cloud. Repeated CH and \CHP\ observations, performed using the star
\zper\ as reference, confirm the reality of rapid and 
large amplitude variations of $N$(CH) and $N$(\CHP) along the line of sight. 
 
$\bullet$ The results altogether strongly suggest that the recent arrival of
HD~34078 near the southern edge of a molecular cloud has given rise
to a shell of dense gas at the interface between the stellar wind and
\H2\ gas, the latter material belonging either to the distorted boundary
of the ambient cloud or to a small foreground clump, as suggested by 
\citet{franceal04}. The location of this shell relative to the star is 
consistent
with constraints derived from earlier modelling of \H2\ excitation. By
comparing the geometrical characteristics of the IR arc detected 
by \citet{franceal07} and the velocity field inferred from our CO
or optical observations to predictions of steady-state bow shock models, 
we find that the latter are inconsistent with the observed
properties. Therefore, we may be seeing this region at an early phase
of the wind/cloud interaction, with a dense layer formed at the interface
but no stationary flow yet established. 

$\bullet$ We propose that the large relative CH and \CHP\ abundances
originate from significant overproduction of \CHP\ in the dense shell, due to
the presence of a strong C-shock and/or of mixing of warm
ionised gas and highly excited molecular material at the wind/cloud 
interface. The pronounced, 
correlated CH and \CHP\ variations would then reflect marked chemical 
structure in the dense shell, possibly resulting from instabilities 
occuring at the interface. 

$\bullet$ No variations of $N$(\H2, $J=0$) have been found at a level of
5\% (3$\sigma$ limit), extending the result obtained in B05 to a time
interval of 4 years, or 110 AU. In the scenario that we propose, $J=0$ 
\H2\ molecules are mainly located beyond the dense shell. This indicates that 
beyond the photodissociation and photoionisation fronts, where the molecular
material is not yet affected by the interaction with HD~34078, no marked
small-scale scale structure is present and that the bulk of the mass is
distributed relatively uniformly within the cloud. \\

Let us now discuss some prospects about the time evolution of HD~34078's 
environment, assuming the scenario sketched in this
paper is roughly correct.
One may in particular wonder whether the future evolution 
can induce significant observable changes in the coming
years. Using the estimates for the observed standoff distance and
its expected steady-state value given in Appendix B, 
one can get a lower limit for the time 
needed for a stationary flow to establish, $(R_{0, \mbox{obs.}} - R_{0,\mbox{th.}})
/V_s \simeq R_{0, \mbox{obs.}}/V_s \simeq 350$ yr. We therefore expect very 
little change in the morphology of 
the mid-IR or CO emission. The only appreciable evolution would involve
the motion of the star relative to the IR or CO peak (1~\arcsec\ in 23
yrs), since the shell is supposed to remain nearly static in the near
future.

However, during its evolution towards a stationary dynamical regime, 
the velocity field must undergo a drastic variation to reach the 
steady-state solution. CO mm observations, which provide
excellent spectral resolution, might thus reveal significant velocity
changes. A higher spatial resolution map of CO emission would be useful in
this regard and of great help to better understand the kinematics
underlying the remarkable pattern observed for line profiles in Fig. 1.
In the next decade, ALMA will offer excellent opportunities for such 
observations. Numerical simulations of the
time evolution of the wind/cloud interface (which, to our knowledge are
not available) would also be very
useful to indicate how the evolution of the velocity field will proceed
in the early phase. 

In the above reasoning, we implicitly assumed that the dense shell will
smoothly evolve towards the steady-state solution but this is in no way
evident. If indeed instabilities develop efficiently at the interface
(as suggested by CH and \CHP\ variations), the cloud may simply be
gradually destroyed as the star moves. Such a picture would be
consistent with the suggestion by \citet{herbig58} that the absence of IS
material south of the star is due to ``clearing'' along the path
followed by HD~34078 in the past (the clump involved in the second
scenario might then simply be a fragment of the initial cloud in the
process of photoevaporation).

Focusing now on the present state of HD~34078's environment, we note
that it represents a remarkable PDR and shock for which many observational
constraints are or might be available, thanks to the presence of a
background UV-bright star. Geometrical parameters are now well
determined and physical conditions in the ambient cloud relatively 
well constrained (sensitive CO emission observations further away 
from HD~34078 would allow us to better characterize them and verify 
that the peculiar excitation conditions determined in Sect. 2.2 are 
specific to the immediate vicinity of the star). Then, the HD~34078 PDR 
may be used as a reference to test our understanding of various 
physical and chemical processes occuring elsewhere at cloud interfaces 
subject to less extreme conditions. 

\begin{acknowledgements}
We warmly thank observers at OHP and S. Ilovaisky in particular for spectra
taken there with ELODIE and SOPHIE as well as J. Krelowski, G. Galazutdinov,
F. Musaev and A. Bondar for collecting the Terskol, BOAO and Calar Alto
spectra. We are also indebted to M. Gerin and L. Pagani for their
assistance in the preparation of IRAM observations, to J.M. D\'esert
for help in the FUSE data reduction, and to K. France for providing the 
{\it Spitzer} 24~$\mu$m map. Finally, we are grateful to S. Cabrit, 
F. Martins and J. Zorec for several helpful discussions.
\end{acknowledgements}

\bibliographystyle{aa} 
\bibliography{MIS} 

\begin{thebibliography}{52}
\expandafter\ifx\csname natexlab\endcsname\relax\def\natexlab#1{#1}\fi

\bibitem[{{Allen}(1994)}]{allen94}
{Allen}, M.~M. 1994, \apj, 424, 754

\bibitem[{{Baranne} {et~al.}(1996){Baranne}, {Queloz}, {Mayor}, {Adrianzyk},
  {Knispel}, {Kohler}, {Lacroix}, {Meunier}, {Rimbaud}, \& {Vin}}]{baranneal96}
{Baranne}, A., {Queloz}, D., {Mayor}, M., {et~al.} 1996, \aaps, 119, 373

\bibitem[{{Bertoldi} \& {Draine}(1996)}]{bertoldial96}
{Bertoldi}, F. \& {Draine}, B.~T. 1996, \apj, 458, 222

\bibitem[{{Boiss{\'e}} {et~al.}(2005){Boiss{\'e}}, {Le Petit}, {Rollinde},
  {Roueff}, {Pineau des For{\^e}ts}, {Andersson}, {Gry}, \&
  {Felenbok}}]{boisseal05}
{Boiss{\'e}}, P., {Le Petit}, F., {Rollinde}, E., {et~al.} 2005, \aap, 429,
  509, {(BO5)}

\bibitem[{{Bouchy} \& {The Sophie Team}(2006)}]{bouchyal06}
{Bouchy}, F. \& {The Sophie Team}. 2006, in Tenth Anniversary of 51 Peg-b:
  Status of and prospects for hot Jupiter studies, ed. L.~{Arnold},
  F.~{Bouchy}, \& C.~{Moutou}, 319--325

\bibitem[{{Crane} {et~al.}(1991){Crane}, {Hegyi}, \& {Lambert}}]{craneal91}
{Crane}, P., {Hegyi}, D.~J., \& {Lambert}, D.~L. 1991, \apj, 378, 181

\bibitem[{{Crane} {et~al.}(1995){Crane}, {Lambert}, \& {Sheffer}}]{craneal95}
{Crane}, P., {Lambert}, D.~L., \& {Sheffer}, Y. 1995, \apjs, 99, 107

\bibitem[{{Crawford}(1995)}]{crawford95}
{Crawford}, I.~A. 1995, \mnras, 277, 458

\bibitem[{{Crawford}(2003)}]{crawford03}
{Crawford}, I.~A. 2003, \apss, 285, 661

\bibitem[{{Deshpande}(2007)}]{deshpande07}
{Deshpande}, A.~A. 2007, in Astronomical Society of the Pacific Conference
  Series, Vol. 365, SINS - Small Ionized and Neutral Structures in the Diffuse
  Interstellar Medium, ed. M.~{Haverkorn} \& W.~M. {Goss}, 105--+

\bibitem[{{Duley} {et~al.}(1992){Duley}, {Hartquist}, {Sternberg},
  {Wagenblast}, \& {Williams}}]{duleyal92}
{Duley}, W.~W., {Hartquist}, T.~W., {Sternberg}, A., {Wagenblast}, R., \&
  {Williams}, D.~A. 1992, \mnras, 255, 463

\bibitem[{{Falgarone} {et~al.}(1998){Falgarone}, {Panis}, {Heithausen},
  {Perault}, {Stutzki}, {Puget}, \& {Bensch}}]{falgaroneal98}
{Falgarone}, E., {Panis}, J.-F., {Heithausen}, A., {et~al.} 1998, \aap, 331,
  669

\bibitem[{{Federman} {et~al.}(1994){Federman}, {Strom}, {Lambert}, {Cardelli},
  {Smith}, \& {Joseph}}]{federmanal94}
{Federman}, S.~R., {Strom}, C.~J., {Lambert}, D.~L., {et~al.} 1994, \apj, 424,
  772

\bibitem[{{Flower} \& {Pineau des For\^ets}(1998)}]{floweral98}
{Flower}, D.~R. \& {Pineau des For\^ets}, G. 1998, \mnras, 297, 1182

\bibitem[{{Frail} {et~al.}(1994){Frail}, {Weisberg}, {Cordes}, \&
  {Mathers}}]{frailal94}
{Frail}, D.~A., {Weisberg}, J.~M., {Cordes}, J.~M., \& {Mathers}, C. 1994,
  \apj, 436, 144

\bibitem[{{France} {et~al.}(2004){France}, {McCandliss}, {Burgh}, \&
  {Feldman}}]{franceal04}
{France}, K., {McCandliss}, S.~R., {Burgh}, E.~B., \& {Feldman}, P.~D. 2004,
  \apj, 616, 257

\bibitem[{{France} {et~al.}(2007){France}, {McCandliss}, \&
  {Lupu}}]{franceal07}
{France}, K., {McCandliss}, S.~R., \& {Lupu}, R.~E. 2007, \apj, 655, 920

\bibitem[{{Galazutdinov} {et~al.}(2005){Galazutdinov}, {Han}, \&
  {Kre{\l}owski}}]{galazutdinoval05}
{Galazutdinov}, G.~A., {Han}, I., \& {Kre{\l}owski}, J. 2005, \apj, 629, 299

\bibitem[{{Heiles}(1997)}]{heiles97}
{Heiles}, C. 1997, \apj, 481, 193

\bibitem[{{Herbig}(1958)}]{herbig58}
{Herbig}, G.~H. 1958, \pasp, 70, 468

\bibitem[{{Hilton} \& {Lahulla}(1995)}]{hiltonal95}
{Hilton}, J. \& {Lahulla}, J.~F. 1995, \aaps, 113, 325

\bibitem[{{Hily-Blant} \& {Falgarone}(2007)}]{hilyblantal07}
{Hily-Blant}, P. \& {Falgarone}, E. 2007, \aap, 469, 173

\bibitem[{{Hoogerwerf} {et~al.}(2001){Hoogerwerf}, {de Bruijne}, \& {de
  Zeeuw}}]{hoogerwerfal01}
{Hoogerwerf}, R., {de Bruijne}, J.~H.~J., \& {de Zeeuw}, P.~T. 2001, \aap, 365,
  49

\bibitem[{{Lambert} \& {Danks}(1986)}]{lambertal86}
{Lambert}, D.~L. \& {Danks}, A.~C. 1986, \apj, 303, 401

\bibitem[{{Lauroesch}(2007)}]{lauroesch07}
{Lauroesch}, J.~T. 2007, in Astronomical Society of the Pacific Conference
  Series, Vol. 365, SINS - Small Ionized and Neutral Structures in the Diffuse
  Interstellar Medium, ed. M.~{Haverkorn} \& W.~M. {Goss}, 40--+

\bibitem[{{Lesaffre} {et~al.}(2007){Lesaffre}, {Gerin}, \&
  {Hennebelle}}]{lesaffreal07}
{Lesaffre}, P., {Gerin}, M., \& {Hennebelle}, P. 2007, \aap, 469, 949

\bibitem[{{Lien}(1984)}]{lien84}
{Lien}, D.~J. 1984, \apj, 284, 578

\bibitem[{{Liszt} \& {Lucas}(2000)}]{lisztal00}
{Liszt}, H. \& {Lucas}, R. 2000, \aap, 355, 333

\bibitem[{{Liszt} \& {Lucas}(1998)}]{lisztal98}
{Liszt}, H.~S. \& {Lucas}, R. 1998, \aap, 339, 561

\bibitem[{{Mac Low} {et~al.}(1991){Mac Low}, {van Buren}, {Wood}, \&
  {Churchwell}}]{mclowal91}
{Mac Low}, M.-M., {van Buren}, D., {Wood}, D.~O.~S., \& {Churchwell}, E. 1991,
  \apj, 369, 395

\bibitem[{{Marchenko} {et~al.}(1998){Marchenko}, {Moffat}, {van der Hucht},
  {Seggewiss}, {Schrijver}, {Stenholm}, {Lundstrom}, {Setia Gunawan},
  {Sutantyo}, {van den Heuvel}, {de Cuyper}, \& {Gomez}}]{marchenkoal98}
{Marchenko}, S.~V., {Moffat}, A.~F.~J., {van der Hucht}, K.~A., {et~al.} 1998,
  \aap, 331, 1022

\bibitem[{{Martins} {et~al.}(2005){Martins}, {Schaerer}, {Hillier},
  {Meynadier}, {Heydari-Malayeri}, \& {Walborn}}]{martinsal05}
{Martins}, F., {Schaerer}, D., {Hillier}, D.~J., {et~al.} 2005, \aap, 441, 735

\bibitem[{{McLachlan} \& {Nandy}(1984)}]{mclachlanal84}
{McLachlan}, A. \& {Nandy}, K. 1984, \mnras, 207, 355

\bibitem[{{Meyer} {et~al.}(2001){Meyer}, {Lauroesch}, {Sofia}, {Draine}, \&
  {Bertoldi}}]{meyeral01}
{Meyer}, D.~M., {Lauroesch}, J.~T., {Sofia}, U.~J., {Draine}, B.~T., \&
  {Bertoldi}, F. 2001, \apjl, 553, L59

\bibitem[{{Moore} \& {Marscher}(1995)}]{mooreal95}
{Moore}, E.~M. \& {Marscher}, A.~P. 1995, \apj, 452, 671

\bibitem[{{Pellerin} {et~al.}(2002){Pellerin}, {Fullerton}, {Robert}, {Howk},
  {Hutchings}, {Walborn}, {Bianchi}, {Crowther}, \& {Sonneborn}}]{pellerinal02}
{Pellerin}, A., {Fullerton}, A.~W., {Robert}, C., {et~al.} 2002, \apjs, 143,
  159

\bibitem[{{Pety} {et~al.}(2006){Pety}, {Gueth}, {Guilloteau}, \&
  {Dutrey}}]{petyal06}
{Pety}, J., {Gueth}, F., {Guilloteau}, S., \& {Dutrey}, A. 2006, \aap, 458, 841

\bibitem[{{Pety} {et~al.}(2008){Pety}, {Lucas}, \& {Liszt}}]{petyal08}
{Pety}, J., {Lucas}, R., \& {Liszt}, H.~S. 2008, ArXiv e-prints, 808

\bibitem[{{Pfeiffer} {et~al.}(1998){Pfeiffer}, {Frank}, {Baumueller},
  {Fuhrmann}, \& {Gehren}}]{pfeifferal98}
{Pfeiffer}, M.~J., {Frank}, C., {Baumueller}, D., {Fuhrmann}, K., \& {Gehren},
  T. 1998, \aaps, 130, 381

\bibitem[{{Pichon} {et~al.}(2001){Pichon}, {Vergely}, {Rollinde}, {Colombi}, \&
  {Petitjean}}]{pichonal01}
{Pichon}, C., {Vergely}, J.~L., {Rollinde}, E., {Colombi}, S., \& {Petitjean},
  P. 2001, \mnras, 326, 597

\bibitem[{{Pineau des For\^ets} {et~al.}(1986){Pineau des For\^ets}, {Flower},
  {Hartquist}, \& {Dalgarno}}]{pineaual86}
{Pineau des For\^ets}, G., {Flower}, D.~R., {Hartquist}, T.~W., \& {Dalgarno},
  A. 1986, \mnras, 220, 801

\bibitem[{{Rollinde} {et~al.}(2003){Rollinde}, {Boiss{\'e}}, {Federman}, \&
  {Pan}}]{rollindeal03}
{Rollinde}, E., {Boiss{\'e}}, P., {Federman}, S.~R., \& {Pan}, K. 2003, \aap,
  401, 215, {(R03)}

\bibitem[{{Schuster} {et~al.}(2004){Schuster}, {Boucher}, {Brunswig}, {Carter},
  {Chenu}, {Foullieux}, {Greve}, {John}, {Lazareff}, {Navarro}, {Perrigouard},
  {Pollet}, {Sievers}, {Thum}, \& {Wiesemeyer}}]{schusteral04}
{Schuster}, K.-F., {Boucher}, C., {Brunswig}, W., {et~al.} 2004, \aap, 423,
  1171

\bibitem[{{Sheffer} {et~al.}(2008){Sheffer}, {Rogers}, {Federman}, {Abel},
  {Gredel}, {Lambert}, \& {Shaw}}]{shefferal08}
{Sheffer}, Y., {Rogers}, M., {Federman}, S.~R., {et~al.} 2008, ArXiv e-prints,
  807

\bibitem[{{van Buren} \& {Mac Low}(1992)}]{vanburenal92}
{van Buren}, D. \& {Mac Low}, M.-M. 1992, \apj, 394, 534

\bibitem[{{van Buren} {et~al.}(1990){van Buren}, {Mac Low}, {Wood}, \&
  {Churchwell}}]{vanburenal90}
{van Buren}, D., {Mac Low}, M.-M., {Wood}, D.~O.~S., \& {Churchwell}, E. 1990,
  \apj, 353, 570

\bibitem[{{van Buren} {et~al.}(1995){van Buren}, {Noriega-Crespo}, \&
  {Dgani}}]{vanburenal95}
{van Buren}, D., {Noriega-Crespo}, A., \& {Dgani}, R. 1995, \aj, 110, 2914

\bibitem[{{Weisberg} \& {Stanimirovi{\'c}}(2007)}]{weisbergal07}
{Weisberg}, J.~M. \& {Stanimirovi{\'c}}, S. 2007, in Astronomical Society of
  the Pacific Conference Series, Vol. 365, SINS - Small Ionized and Neutral
  Structures in the Diffuse Interstellar Medium, ed. M.~{Haverkorn} \& W.~M.
  {Goss}, 28--+

\bibitem[{{Welty}(2007)}]{welty07}
{Welty}, D.~E. 2007, \apj, 668, 1012

\bibitem[{{Welty} {et~al.}(2006){Welty}, {Federman}, {Gredel}, {Thorburn}, \&
  {Lambert}}]{weltyal06}
{Welty}, D.~E., {Federman}, S.~R., {Gredel}, R., {Thorburn}, J.~A., \&
  {Lambert}, D.~L. 2006, \apjs, 165, 138

\bibitem[{{Welty} {et~al.}(2008){Welty}, {Simon}, \& {Hobbs}}]{weltyal08}
{Welty}, D.~E., {Simon}, T., \& {Hobbs}, L.~M. 2008, \mnras, 388, 323

\bibitem[{{Wilkin}(1996)}]{wilkin96}
{Wilkin}, F.~P. 1996, \apjl, 459, L31

\end{thebibliography}

\onecolumn

\begin{table*}
  \caption{Observation parameters. The projection center of all the data 
    is: $\alpha_{2000} = 05^h16^m18.15^s$, $\delta_{2000} = 34\deg18'44.3''$.}
  \begin{center}
    {\tiny
      \begin{tabular}{rcrlcccrclccr}
        \hline \hline
        Molecule & Transition & Frequency  & Instrument & \# Pix. & F$_{eff}$ & B$_{eff}$ & Resol.
& Resol. & Int. Time & T${sys}$ & Noise & Obs. date \\
                 &            & GHz        &            &         &         &         & arcsec
& \kms{} & hours           &    K    &  mK   &   \\
        \hline
        \twCO{} & \Jone{} & 115.271202 & 30m/AB100 & 2       & 0.95 & 0.74 & 22 & 0.20 & 0.7
& 240 & 20 & 13 Feb. 2004 \\
        \twCO{} & \Jtwo{} & 230.538000 & 30m/HERA  & 9 & 0.91 & 0.52 & 11 & 0.40 & 0.7
& 300 & 10 & 11-13 Feb. 2004 \\

        \hline
      \end{tabular}}
  \end{center}
\label{t:CO}
\end{table*}

\begin{table}[htb]
 \caption{List of observations and measured CH and \CHP\ equivalent
 widths (in m\AA; uncertainties as estimated in Sect. 3.2.1 are given in
 upper index). 
$^a$\ Resolution near \CHl. $^b$\ \WCHl\ and \WCHPlb\ values from OHP/ELODIE spectra
have been corrected by +6\% and +8\% respectively (see text). $^c$\
 Column densities in units of $10^{14}$ \units{cm}{-2}, measured
 directly from line profiles for high resolution spectra (McDonald) or 
computed from $W$ values using Eqs 1 and 2 for other data.}
\begin{tabular}{|c|c|c|c|c|c|c|c|}
 \hline
       &       &      &       &      &   && \\
 Epoch & Observatory & R$^a$ & $W_{4300}$(CH) [m\AA]& $W_{4232}$(CH$^+$)& $W_{4300}$(CH)& $N$(CH)$^c$& $N$(\CHP)$^c$\\
       &       & & \multicolumn{2}{c|}{HD~34078}&   \zper& \multicolumn{2}{c|}{HD~34078} \\
 \hline
 2002.73&            &       &  $57.35^{0.92}$    & $47.30^{1.51}$ & ---	       & $0.97^{0.02}$ & $0.69^{0.02}$\\
 2003.15&           &       &  $54.06^{0.76}$   & $44.50^{1.19}$ & $15.05^{0.39}$      & $0.89^{0.01}$& $0.63^{0.02}$\\
2003.71&            &       &  $52.36^{0.93}$    & $46.44^{1.30}$ &$16.22^{0.46}$      & $0.85^{0.02}$ & $0.67^{0.02}$\\
2003.96&            &       &  $55.54^{1.77}$    & ----& $15.26^{0.58}$                & $0.92^{0.03}$&---\\
2004.15& OHP/ELODIE$^b$ & 32,000  &  $51.94^{0.81}$    & $46.66^{1.30}$& $16.22^{0.99}$& $0.84^{0.01}$& $0.67^{0.02}$ \\
2004.88&            &       &  $54.70^{1.20}$    & ----& ----                          & $0.90^{0.02}$&---\\
2005.72&            &       &  $51.30^{0.90}$    & $45.36^{1.40}$& $15.16^{0.64}$      & $0.82^{0.01}$& $0.65^{0.02}$\\
2006.11&            &       &  $48.97^{0.95}$    & $42.01^{1.40}$& ----                & $0.77^{0.01}$& $0.58^{0.02}$\\
\hline
2006.71&    &   & $54.91^{0.32}$&   $44.23^{0.52}$& $15.98^{0.13}$                     & $0.909^{0.005}$& $0.623^{0.007}$\\
2007.14& OHP/SOPHIE  & 75,000   & $51.00^{0.51}$&   $41.30^{0.45}$& $16.01^{0.26}$     & $0.815^{0.008}$& $0.563^{0.006}$ \\
2007.67&    &   & $53.30^{0.36}$&   $45.33^{0.47}$& $16.05^{0.13}$                     & $0.869^{0.006}$& $0.646^{0.007}$ \\
2008.10&    &   & $54.16^{0.30}$&   $48.29^{0.43}$& $16.22^{0.15}$                     & $0.890^{0.005}$& $0.710^{0.006}$ \\
\hline
2003.00&    &   & $55.50^{0.28}$&   $44.78^{0.45}$& $15.80^{0.14}$                     & $0.880^{0.004}$& $0.632^{0.006}$ \\
2004.02&    &   & $52.10^{0.32}$&   $47.00^{0.51}$& $14.90^{0.36}$                     & $0.816^{0.005}$& $0.672^{0.007}$\\
2004.76&    &   & $50.80^{0.69}$&----& $15.70^{0.29}$                                  & $0.816^{0.01}$&---\\
2004.98& McDonald    & 170,000  & $53.10^{0.33}$&   $47.30^{0.57}$& $15.70^{0.18}$     & $0.843^{0.005}$& $0.697^{0.008}$\\
2005.81&    &   & $49.00^{0.26}$&   $44.90^{0.60}$& $16.00^{0.08}$                     & $0.763^{0.004}$& $0.636^{0.009}$ \\
2005.99&    &   & $49.90^{0.22}$&   $43.50^{0.60}$& $16.15^{0.11}$                      & $0.774^{0.003}$ & $0.613^{0.008}$\\
2007.02&    &   & $53.20^{0.40}$&   $42.70^{0.42}$& $15.88^{0.18}$                    & $0.832^{0.006}$ & $0.595^{0.006}$ \\
2008.00&    &   & $54.34^{1.08}$&   $47.88^{1.64}$& $15.74^{0.15}$                     & $0.90^{0.02}$ & $0.68^{0.02}$ \\
\hline
2004.10& BOAO    & 30,000  & $52.50^{0.45}$&  ----& ---                                          & $0.850^{0.007}$ & ---\\
2004.18& BOAO-Terskol    & 30,000 - 120,000  & $50.67^{0.35}$&  ----&---                                 & $0.807^{0.006}$ &---\\
2006.77& BOAO  & 45,000 & $54.60^{0.62}$&   $44.40^{0.60}$& ---                                 & $0.90^{0.01}$ & $0.627^{0.008}$ \\
\hline
2004.65& Terskol& 120,000&	 $55.00^{0.45}$& ----& ---                             & $0.911^{0.007}$ &---\\
\hline
2007.18& Calar Alto& 40,000 &	 $52.15^{0.88}$& $41.80^{0.95}$& ---                            & $0.84^{0.02}$ & $0.57^{0.01}$ \\
\hline
\end{tabular}
\label{t:obs}
\end{table}

\twocolumn

\appendix
\section{The contribution of the IC~405 nebula to HD~34078 spectra}
\label{s:nebula}

\citet{franceal04} observed the diffuse emission from IC~405 at four
positions (their Pos1 to Pos4), two of which (Pos1 and Pos2)
are located close to the HD~34078 position
(75~\arcsec\ offset, E and W respectively).
Thus, diffuse emission is certainly present towards
HD~34078 itself and may contribute significantly to the flux collected
in the FUSE apertures. To estimate this contribution, we retrieved from the
FUSE database the spectra obtained at Pos1 to Pos4 \citep[cf. Fig. 1
from][]{franceal04}. Pos2 is the brightest region; around 1050~\AA, the
surface brightness is about  30\% larger than at Pos1, indicating
spatial variations of this emission. Since Pos1 and Pos2 are
symetrically located
with respect to HD~34078, a first order estimate of the diffuse flux
towards the star is the average of Pos1 and
Pos2 spectra. We thus estimate that the diffuse flux received in the
LWRS aperture is about 7\% of the HD~34078 flux around $\lambda = 1050$
~\AA\ \citep[this fraction decreases with wavelength since the diffuse to
stellar flux ratio gets lower at longer wavelengths, as discussed by][]{franceal04}.
If the surface brightness is locally uniform over the LWRS
aperture centred on HD~34078's position, the contribution of diffuse
emission scales linearly with aperture size and should be about 11 times
lower in MDRS spectra. Thus, while LWRS spectra of HD~34078 are
significantly affected by diffuse emission, the MDRS spectrum should be
essentially free of such ``pollution'', except possibly at the shortest wavelengths.

To estimate the impact on HD~34078's LWRS spectra of the contamination
by diffuse emission, one needs to examine the spectrum of the
latter \citep[the Pos1 spectrum is displayed in
Fig. 7 from][]{franceal04}.
 Although its S/N ratio is limited, it is clear that it differs
from that of HD~34078 in two respects. First, narrow lines
appear to be fainter and shallower (only the strongest lines are detected); this
is due, at least in part, to the lower effective resolution implied by 
the extented
nature of the source. 
Second, broad \H2\ lines are narrower, indicating that the average
pathlength of scattered photons through the molecular gas is smaller 
than that followed by direct HD~34078 photons. These two properties of 
diffuse emission qualitatively explain the
peculiarities of the 8$^{\rm th}$ spectrum and indeed, one finds that by 
combining it 
and our estimate of the diffuse emission spectrum towards
HD~34078, it is possible to reproduce spectra 1 to 7 fairly well. 
We conclude that the apparent changes in the 8$^{\rm th}$ spectrum can be
attributed mainly to a smaller contribution of diffuse emission due
to the use of MRDS instead of LWRS in the earlier spectra. 

\section{The steady-state bow shock model confronted to observations}
Extensive work has been performed to describe the geometry and velocity
structure of steady-state bow shocks
\citep[e.g.][]{vanburenal90,mclowal91,wilkin96}
 and in the thin-shell limit, 
there are simple analytical predictions that can be directly compared to
observations. 
 
\subsection{Shape and radius of the shell}

Assuming that the IR arc detected by \citet{franceal07} corresponds
to a steady-state bow shock viewed in projection onto the sky, we can 
first check whether the apparent geometrical properties are consistent
with the predicted ones. 
At first sight, the distance between the star and the apex of the bow 
shock looks too small as compared to its radius of curvature on the 
{\it Spitzer} image but we have to 
account for projection effects which somewhat influence the appearence 
of the arc. HD~34078's tangential and 
radial components ($V_{\mbox{t}}$ and $V_{\mbox{r}}$ respectively) are well constrained 
by observations; we adopt $V_{\mbox{t}} = 100\, \kms$ and $V_{\mbox{r}}
= + 59\, \kms$ (in the LSR system; we remeasured the latter value from our 
own visible spectra). The velocity vector is then inclined by an angle
$\theta = 30\degr$ with respect to the plane of the sky (\Fig{cartoon}).

To estimate the impact of projection effects on the $h/R_{c,p}$
ratio, where $h$ is the apparent standoff distance \citep[$h/D \simeq 10 - 20
~\arcsec$\ after][ we adopt $h/D=15~\arcsec$ with $D = 530$~pc, 
the distance to HD~34078]{franceal07} and 
$R_{c,p}$ is the radius of curvature of the arc as seen in projection on
the sky ($R_{c,p}/D \simeq$ 37~\arcsec), we approximate the bow shock
geometry by a paraboloid \citep{vanburenal90,wilkin96} and find
after some algebra that 

\begin{equation}
h = \frac {3+\cos^2\theta}{4\cos\theta} R_0
\end{equation}
 and 
\begin{equation}
R_{c,p} = \frac {3}{2 \cos\theta} R_0,
\end{equation}

\noindent where $R_0$ is the distance 
between the star ($S$) and the apex ($A$) of the shock.
With $\theta = 30\degr$, we get $h=1.08 R_0$ and $h/R_{c,p} = 0.62$
while the estimate inferred from the IR map is $h/R_{c,p} = 0.40$
(the upper limit is 0.54 for $h = 20~\arcsec$). 
Thus, projection effects appear to be insufficient to explain the 
relative large radius of curvature of the arc. Moreover, it seems
difficult to explain in this model why the arc does not extend further 
southward \citep[cf. Fig. 5 in][]{franceal07}. Given that projection
effects are in the end very limited, the standoff distance can be
estimated to be about 0.04~pc.

Although the observed arc shape is not well fitted by the model
prediction, the IR data can nevertheless be used to get a rough estimate 
of the distance between the star and the point where 
the line of sight intersects the shell ($L_2$ in \Fig{cartoon}), 
by making the reasonable
assumption that the latter is axially symmetric. From the 24~$\mu$m
map, we estimate that $d(S,L_2)/D \simeq 40~\arcsec$, implying $d
=0.10$~pc. Is this value compatible with the UV flux necessary to explain
the amount of highly excited \H2\ ? B05 found that a radiation
field about $10^4$ larger than that in the local ISM is required, which
is obtained at 0.2 pc from the
HD~34078, a value in reasonable agreement with our estimate for
$d(S,L_2)$. 

\subsection{Momentum balance and ambient density}
In steady-state, the standoff distance, $R_0$, is set by a momentum balance
equation (Eq. 2 in van Buren et al. 1990; Eq. 1 in Wilkin 1996) which is  

\begin{equation}
R_0 =  \sqrt{\frac {\dot{m}_{\star} V_{\mbox{w}}}{4\pi \rho_{\mbox{a}} V^{2}_{\star}}}
\end{equation}

\noindent where 
 $\dot{m}_{\star}$ is the mass loss rate, $V_s$ the terminal wind
velocity \citep[$10^{-9.5}$ M$_{\sun}$ yr$^{-1}$ and 800 \kms\
respectively after][]{martinsal05}, $\rho_{\mbox{a}}$ the mass 
density of the ambient
medium and $V_s$ the star velocity (116 \kms\ with the above values for $V_{\mbox{t}}$
and $V_{\mbox{r}}$). 
The ambient H number density can be estimated either from \ion{C}{i}$^{(*,**)}$ 
($n = 700\, \cmt$; B05) or C$_2$ absorption lines 
\citep[$n = 300\, \cmt$;][]{federmanal94}, assuming the latter are not significantly
contaminated by the dense shell. Adopting $n = 500$~\cmt, we get a 
``theoretical'' value of $R_0 = 3.5 \times 10^{-4}$ pc while 
observations indicate $R_0 \simeq 0.04$ pc, in marked
disagreement. In other words, an ambient density as low as of a 
few $10^{-2}$~\cmt\ would be required for a steady-state bow shock to be
at the observed distance, which is highly unrealistic for molecular-rich 
gas. Radiative pressure from stellar photons might help to maintain
the shell at a distance larger than expected on the basis of the wind
pressure alone. One can obtain easily an upper limit for the radiative
to wind pressure ratio, $P_{\mbox{rad}}/P_{\mbox{w}}$, by assuming that all photons
impinging on the shell are absorbed: 
$P_{\mbox{rad}}/P_{\mbox{w}} = L/(\dot{m}_{\star} c V_{\mbox{w}})$. This ratio 
is of the order of $10^{-3}$ for HD~34078; radiation pressure is
therefore negligible here.

Since $P_{\mbox{w}}$ scales linearly with $\dot{m}_{\star}$, one
may wonder whether the mass loss rate has been underestimated. Prior to 
the study by \citet{martinsal05}, the adopted value for HD~34078 was 
$10^{-6.6}$ M$_{\sun}$ yr$^{-1}$ (i.e. larger by a factor of 800 than the present estimate !) 
and even with this much higher rate, the required ambient density would 
amount only to $n \simeq 20\, \cmt$. The much lower recent mass loss 
estimate is based on the availability of UV lines (\ion{C}{iv}~$\lambda$1550 
mainly) which better probe 
weak winds; the uncertainty on the revised value is estimated to be of 
a factor of about 3 (F. Martins; private communication). Then,
$\dot{m}_{\star}$ cannot have been underestimated by a factor large enough
to explain the discrepancy between the observed and theoretical
steady-state $R_0$ values.

Stationary bow shock models also provide specific predictions for the mass
surface density or equivalently the column density $N{\mbox{(H)}}$ of swept-up
material trapped in the bow shock (cf Eq. 7 from van Buren et al.
1990 or Eq. 12 from Wilkin 1996). $N{\mbox{(H)}}$ scales as $\dot{m}^{1/2}_{\star}V^{1/2}_{\mbox{w}} V^{-1}_{\mbox{s}} n^{1/2}$ and with the values quoted above, we get
$N{\mbox{(H)}}=6.0 \times 10^{17} \cmd$. This prediction is to be compared to
the H column density in the hot PDR component of B05: $2.7 \times
10^{19}\, \cmd$, including \H2\ only (their Tab. 4). Note that their 
exceedingly large predicted value for the \HI\ column density was due to the
assumption of steady-state equilibrium for the photodissociation of
\H2; in our scenario, this assumption is no longer realistic.
The observed value corresponds to the dense material along the line of 
sight to HD~34078 (i.e. located at $L_2$ in \Fig{cartoon}) while the 
``theoretical'' one refers to the apex position ($A$ in \Fig{cartoon}).
This does not make a large difference however since Wilkin's results
(his Fig. 4) indicate that the surface density normal to the bow
shock varies slowly with position away from the apex. One should also
consider that in the geometry of \Fig{cartoon}, the shell is not crossed 
normally by the line of sight but with an inclication angle $\varphi$,
but this involves a factor of at most a few. Obviously, this cannot
explain the large discrepancy between the two values above. 

In the 24~$\mu$m map, the arc defines a roughly hemispherical cavity (with
a radius of about $R_{c,p}$) and one can easily get another estimate for the
column density in the compressed shell by simply assuming that material
from the ambient cloud (i.e. with $n=500$~\cmt) initially filling this
cavity has been swept by the stellar wind to form the present
shell. This leads to a column density of $N = n R_{c,p}/3$ and
interestingly, this expression provides a value, $N \simeq 5 \times
10^{19}$~\cmd, comparable to the
estimate of B05 for the hot PDR component.

\subsection{Velocity field of the compressed gas}
\label{s:velocity}
Another way to assess whether a steady-state bow-shock model is consistent
with our data is to compare the observed and predicted velocity fields.
Since \co21\ emission traces dense gas within the shell of compressed gas,
our CO data can be used to constrain the velocity field around HD~34078.
One may wonder in particular, whether the double-peaked profiles with their
remarkable symmetry properties (Fig. 1) simply arise from the fact that the
IRAM beam intersects the paraboloidal wind/cloud interface twice. To
compute a "synthetic" \co21\ emission map, we adapted the model developed
by \citet{petyal06} to describe the outflow around HH30. We relied on
the analytical expressions provided by \citet{wilkin96} for the geometry,
velocity field and mass surface density \citep[see also][]{vanburenal92},
with the parameter values considered above. For practical reasons, the
thickness of the boundary layer has been assumed to be $R_0$/20 (while it
is zero in Wilkin's model) and the absolute surface mass density has been
scaled so as to reproduce the observed intensity. The underlying assumption
is that the medium is optically thin which is reasonable given the
strength of the CO emission. The resulting signal model was convolved with
the IRAM-30m 230~GHz beam, an important step given the beam size relative
to the source extent. Finally, in order to assess whether a specific model
is acceptable or not, we compared the synthetized and observed maps of
spectra as well as position-velocity and channel maps.

Several difficulties arised when confronting the model with observations.
First, the extent of the \co21\ emission tends to be too limited if one
adopts $R_0/D$ = 15 arcsec. This is related to the fact that the expected
radius of curvature is too small for such a $R_0$ value, as compared to the
observed one (cf above).  Since here we are mainly interested in the
velocity field, we simply adjusted $R_0$ so as to match the observed extent
of the emission. Second, the velocity range over which \co21\ emission
appears in the model is much larger than the observed one. Indeed, the
velocity field of the gas scales linearly with $V_s$: in particular, the
typical separation $\Delta V$ between the two CO emission peaks near
HD~34078's position should be of the order of 0.7 $V_s$ while we observe
only 0.02 $V_s$ (corresponding to about 2 \kms). Artificially modifying the
star's velocity to the former value allows us to qualitatively reproduce the
emission properties close to the star. However, the double-peaked
character of the model profiles tend to be less pronounced that in the
observed ones. Other velocity/density distributions might be considered to get
a better fit, but clearly, only higher spatial resolution observations
would allow us to get unambiguous constraints on such models.

Another constraint can be obtained from optical absorption lines arising towards
HD~34078. With the parameters quoted above, the expected velocity of
the gas from the shell along the sightline to HD~34078 (i.e. at $L_2$)
is $\vlsr \simeq -15$ \kms. Such a shift between highly 
excited \H2\ lines (tracing the dense layer) and absorption from
species located beyond the shell would be easily detectable, but B05 
failed to find any significant velocity difference between the two
components. 

To summarize, the dense shell at the stellar wind/molecular cloud
interface (with a density of about $n \simeq 10^4\, \cmt$ and column density 
$N{\mbox{(H)}} \simeq 3 \times 10^{19}\, \cmd$, thus corresponding to a thickness 
of $\simeq 10^{-3}$ pc) is located at a distance of the star that is consistent
with the excitation of \H2\ but it does not display the properties 
expected for a steady-state bow shock: 
i) the arc is not curved enough;  ii) the shell is too far from the star
for the momentum balance to be satisfied; iii) the amount of material 
swept up by the wind is too large
and iv) the velocity field shows
very little deviation from the ambient value. The limited extent and
somewhat irregular geometry of the arc are additional indications against a
steady-state bow shock.
\end{document}